\documentstyle[charm2000]{article}
\input psfig

\begin{document}

\title{
\begin{flushright}
PRINCETON/HEP/95-6 \\
hep-ph/9508415 \\
\end{flushright}
\large\bf An Overview of $D^0\bar{D}^0$ Mixing Search Techniques: Current
Status and Future Prospects~\thanks{Presented at the
$\tau$-charm Factory Workshop, Argonne National Laboratory,
June 20-23, 1995.}}

\author{\large{Tiehui (Ted) Liu} 		
     \\ \\ {\sl Department of Physics, Princeton University,
Princeton, NJ 08544}	
         }  

\maketitle


\begin{abstract}

The search for $D^0\bar{D}^0$
mixing may carry a large discovery potential for new
physics since the $D^0\bar{D}^0$
mixing rate is expected to be small in the Standard
Model.  The past decade has seen significant experimental progress in
sensitivity.
This paper discusses the techniques,
current experimental status, and future prospects for the mixing
search. Some new ideas, applicable to future mixing searches,
are introduced. In this paper, the importance of separately measuring
the decay rate difference and the mass difference of the
two CP eigenstates (in order to
observe New Physics) has been emphasized, since the
theoretical calculations for long distance effects are still
plagued by large uncertainties.

\end{abstract}

\section{Introduction}

Particle-antiparticle mixing has always been of fundamental
importance in testing the Standard Model and constraining new
physics.
This is because mixing is responsible for the small mass
differences between the mass eigenstates of neutral mesons. Being
a flavor changing neutral current (FCNC) process,
it often involves heavy quarks in loops. Such higher order processes are of
great interest since the amplitudes are sensitive to any
weakly-coupling quark flavor running around the loop.
Historically, $K^0\bar{K}^0$ mixing is the rare (FCNC) process that has been
experimentally examined in the greatest detail. It has been amply demonstrated
that in spite of many inherent uncertainties of strong
interaction physics, the Standard Model predicts the correct phenomenology of
the $K^0\bar{K}^0$ mixing. In fact, based on the calculation of the $K_{L}$ -
$K_{S}$
mass difference, Gaillard and Lee~\cite{Gaillard} were able to estimate the
value of the charm
quark mass before the discovery of charm. Moreover,
$B^0\bar{B}^0$ mixing gave the first indication of a large top
quark mass.

Although $D^0\bar{D}^0$ mixing is very similar to
$K^0\bar{K}^0$ and $B^0\bar{B}^0$ mixing,
as all are FCNC processes, there are significant differences which make
$D^0\bar{D}^0$ mixing a possible unique place to explore
new physics. Roughly speaking,
in the case of $K$ and $B$ FCNC processes, the appearance of
the top quark in the internal loop with $m_{t} > M_{W} >> m_{c},m_{u}$
removes the GIM~\cite{GIM} suppression, making $K$ and $B$ decays a nice place
to test FCNC transitions and to study the physics of the top.
In the case of $D$ FCNC processes, the FCNC are much stronger suppressed
because the down-type quarks ($d$, $s$ and $b$) with $m_{d}$, $m_{s}$, $m_{b}$
$<< M_{W}$ enter the internal loops and the GIM mechanism is much
more effective~\cite{Buras}.
Therefore the $D^0\overline{D}^0$ mixing rate
is expected to be small in the Standard Model,
which means
the mixing search may carry a large potential for discovery of new physics.
There are many extensions
of the Standard Model which allow $D^0\overline{D}^0$
mixing (the mass difference between the two CP eigenstates) to be
significantly larger
than the Standard Model prediction (for example, see~\cite{Babu}
to~\cite{Nir}).
Recent reviews on FCNC processes in $D$ decays
can be found elsewhere~\cite{Burdman,Pakvasa0,Hewett,Schwartz}.
In
general, there could be a large enhancement of the one-loop induced FCNC
processes in $D$ decays with no constraint from limits on FCNC processes
in the $K$ and $B$ systems. Roughly speaking, this is because the couplings
of FCNC to up-type quarks ($u$,$c$,$t$) could be completely different from
those to down-type quarks ($d$,$s$,$b$). Thus one gains independent pieces
of information when searching for FCNC in $D$ decays, compared to what is
learned searching for FCNC in $K$ and $B$ decays.

One can characterize $D^0\bar{D}^0$ mixing in terms of two
dimensionless variables:
$ x={\delta m / \gamma_+}$ and $y={\gamma_- / \gamma_+}$, where
the quantities $\gamma_\pm$ and $ \delta m$ are defined by
$\gamma_\pm = {(\gamma_1\pm \gamma_2) / 2}$ and $ \delta m = m_2 - m_1$
with $m_i,\gamma_i$ $(i=1,2)$ being the masses and decay rates of
the two CP (even and odd) eigenstates.
Assuming a small mixing, namely,
$\delta m, \gamma_- \ll \gamma_+$ or $x,y\ll 1$, we have
${\rm R}_{\rm mixing}= (x^2 + y^2)/2 $.
Mixing can be caused either by $x \neq 0$ (meaning that mixing
is genuinely caused by the $D^0 - \bar{D}^0$ transition) or
by $y \neq 0$ (meaning mixing is caused by the fact that
the fast decaying component quickly disappears, leaving the slow
decaying component which is a mixture of $D^0$ and $\bar{D}^0$).
Theoretical calculations of $D^0\bar{D}^0$ mixing
in the Standard Model are plagued by large uncertainties. While
short distance effects from box diagrams are known~\cite{Gaillard}
to give a negligible contribution ($\sim 10^{-10}$),
the long distance effects from
second-order weak interactions with mesonic intermediate states
may give a much larger contribution. Estimates of
${\rm R}_{\rm mixing}$ from long distance effects range from $10^{-7}$
to $10^{-3}$~\cite{Donoghue}.
It has recently been argued by
Georgi and others that the long
distance contributions are smaller than previously estimated, implying
that cancellations occur between contributions from different
classes of intermediate mesonic states~\cite{Georgi}. While many people
now believe that within the Standard Model
${\rm R}_{\rm mixing}<10^{-7}$~\cite{Burdman,Pakvasa0,Hewett},
others think ${\rm R}_{\rm mixing}$ could be much larger~\cite{Bigi5,Kaeding},
say $10^{-4}$~\cite{Bigi5} (meaning both $x$ and $y$ are above $10^{-3}$).
For example, Bigi~\cite{Bigi5} pointed out
that observing a non-vanishing value for ${\rm R}_{\rm mixing}$ between
$10^{-4}$ and $10^{-3}$ would at present not constitute irrefutable evidence
for New Physics, considering the large uncertainties in the
long distance calculations.
While there is some hope that the uncertainties
can be reduced in the future, as pointed out by Bigi~\cite{Bigi5},
partly through
theoretical efforts and partly through more precise and comprehensive
data (since a more reliable estimate can be obtained from a dispersion
relation involving the measured branching ratios for the channels
common to $D^0$ and $\bar{D^0}$ decays),
one recent paper claims that the hope is rather remote~\cite{Kaeding}.
Speculations abound, but (fortunately) physics is an experimental
science, and only with solid experimental evidence will we be
able to properly address these problems.
As experimentalist, I think the best way is to measure $x$ and $y$
separately, as suggested in~\cite{Charm2000_liu,Liudpf}.
As will be discussed, this is experimentally possible. If we can
measure ${\rm R}_{\rm mixing}$ as well as $y$, then we can in effect measure
$x$. Within the Standard Model, $x$ and $y$ are expected to be at the same
level, although
we do not know exactly at what level as theoretical calculations
for long distance effects (which contribute to both $x$ and $y$) are still
plagued by large uncertainties. We expect New Physics does not affect the
decays in a significant way thus does not contribute to $y$, but only to $x$.
The point I am trying to make here is that
the long distance contribution can be measured, even if it
cannot be calculated in a reliable way;
that is, by measuring $y$ directly. If we can
experimentally confirm that indeed $x >> y$, then we can claim New Physics,
regardless of what theoretical calculations for long distance effects are.
Otherwise, if it turns out $x \sim y$, then mostly likely we are seeing
the Standard Model Physics. Therefore, it is crucial to measure
$y$ in order to understand the size of $x$ within the Standard Model.
This is one of the major points I have been trying to make in
the past~\cite{Charm2000_liu,Liudpf,Thesis} and in this paper.

\begin{figure}[p]
\unitlength 1in
\begin{picture}(6.5,6)(0,0)
\put(-.35,-0.40){\psfig{width=6.26in,height=7.5in,%
file=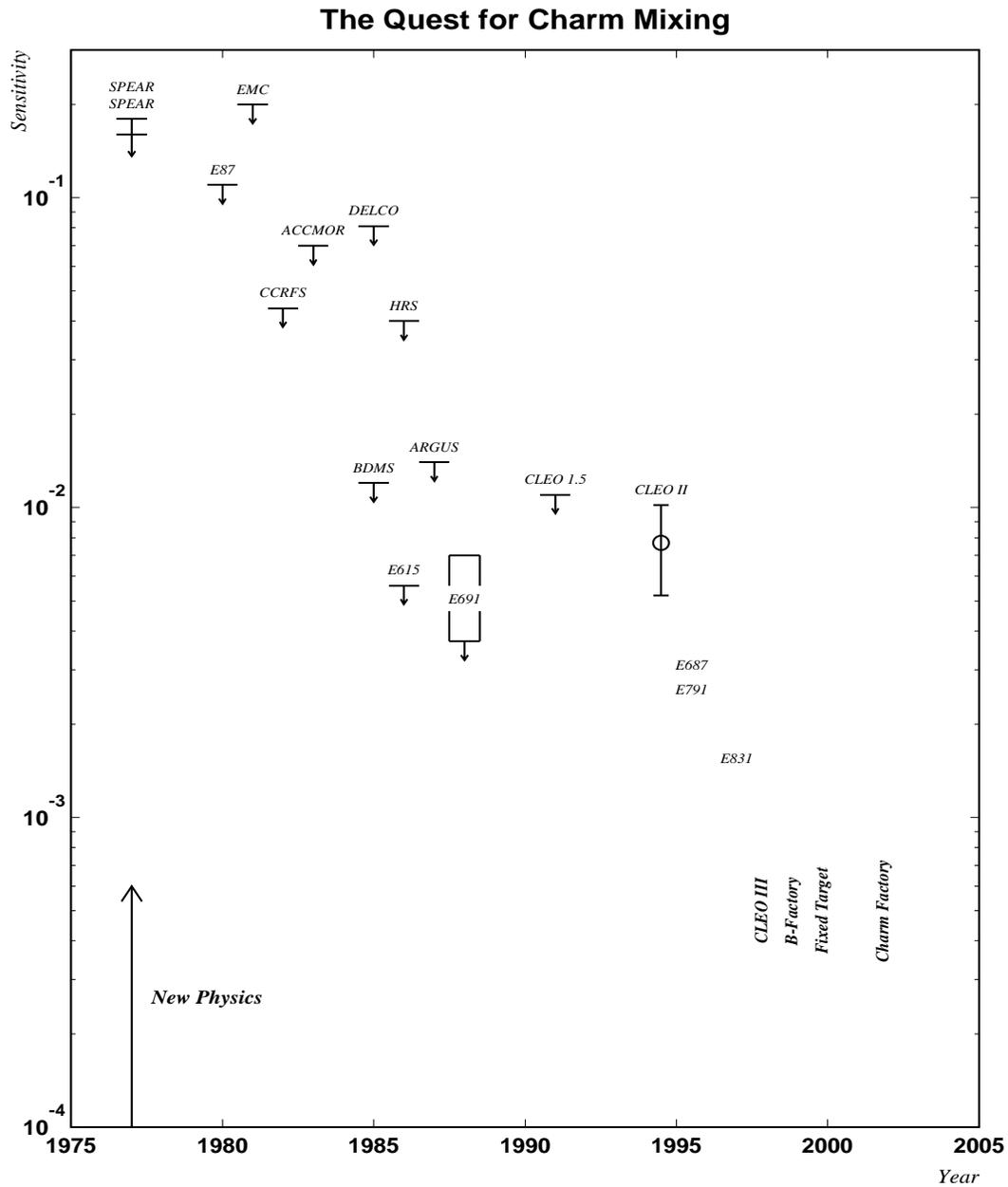}}
\end{picture}
\caption {The history of the quest
for $D^0\bar{D}^0$ mixing.
Note that the range in E691 result
reflects the possible effects of interference between DCSD and mixing,
and the CLEO II signal
could be due to either mixing or DCSD,
or a combination of the two.}
\label{history}
\end{figure}

Motivated by the experience with $K^0\bar{K^0}$ system, experimenters
have been
searching for $D^0\bar{D^0}$ mixing since shortly after the discovery of
$D^0$ meson at SPEAR in 1976, in either hadronic decays $D^0 \to \bar{D^0}
\to K^+\pi^-(X)$~\cite{footnote},
or semileptonic decays $D^0 \to \bar{D^0} \to X^+l^-\nu$.
The past decade has seen significant experimental progress in sensitivity
(from 20\%  to 0.37\%~\cite{SPEAR1}
to~\cite{CLEO15}), as can be seen in
Figure 1.
The search for $D^0\bar{D^0}$ mixing has a long and interesting
history (see Figure~\ref{history}).
In the first few years, people searched for
$D^0 \to K^+\pi^-$ assuming that it would be due to mixing only.
Normally, $D^0$ decays by
Cabibbo favored decay $D^0 \to K^-\pi^+$ and $\bar{D}^0 \to K^+\pi^-$.
A signal for $D^0 \to K^+\pi^-$ could indicate mixing of $D^0 \to \bar{D}^0$.
But it could also indicate a different decay channel, namely,
Doubly Cabibbo Suppressed Decay(DCSD) $D^0 \to K^+\pi^-$,
which is suppressed with respect to the Cabibbo favored decay
by a factor of $tan^4\theta_C \sim 0.3\%$ where
$\theta_C$ is the Cabibbo angle.
As will be discussed, around 1985 there were hints of
$D^0 \to K^+\pi^-\pi^0$ observation, which could be due to
DCSD or mixing. The popular interpretation neglected the possible
DCSD contribution, giving the impression that
$D^0\bar{D^0}$ mixing rate ${\rm R}_{\rm mixing}$
could be of order ${\cal O}(1\%)$.
This engendered much theoretical work to accommodate
the possibly large mixing rate. At that time, the ``theoretical
prejudice'' was that long-distance contributions dominated
and would give a large mixing rate on the order of $1\%$ level.
Later on, fixed target experiments published limits which were not
much larger than the na\"{\i}ve quark model DCSD rate. In light of
these results, the commonly held impression was then that DCSD
was much larger than mixing, and that exploring mixing by means
of hadronic $D^0$ decays had been almost exhausted as a
technique since the ``annoying DCSD background'' would inherently
limit ones ability to observe the interesting physics - $D^0\bar{D^0}$ mixing.
It was believed by many that the signature of mixing
appears only at longer decay times; therefore, it will suffer from DCSD
fluctuation, and
destructive interference could wipe out the signature of mixing.
Since semileptonic decays are not subject to this
``annoying background'', the general consensus was that semileptonic
decays were a better avenue to explore $D^0\bar{D^0}$ mixing.

However, as will be discussed in more detail later,
the commonly believed ``annoying DCSD background'' does not
necessary inherently limit the hadronic method as the
potentially small mixing signature could show up in the
interference term~\cite{Charm2000_liu}.
Moreover, the possible differences between
the resonant substructure in many DCSD and mixing decay modes
could, in principle, be used to distinguish between DCSD
and mixing candidates experimentally~\cite{Charm2000_liu}
(the importance of the
mixing-DCSD interference effect will be more clear here).
Our ability to observe the signature of a potentially
small mixing signal depends on the number of $D^0 \to K^+\pi^-(X)$
events we will have. This means observing $D^0 \to K^+\pi^-(X)$ would
be an important step on the way to observing mixing with this technique.
Recently, CLEO has observed a signal for $D^0\to K^+\pi^-$ (see Figure 2),
and found ${\rm R}$ = ${\cal B}(D^0 \to K^+\pi^-)/$
${\cal B}(D^0 \to K^-\pi^+) \sim 0.8\%$~\cite{Liu}.
Unfortunately, without a precision vertex detector, CLEO is unable to
distinguish a potential mixing signal from DCSD.
If the number of reconstructed charm decays can reach $10^8$ around
the year 2000, that would allow one to reach a new threshold of sensitivity to
$D^0\bar{D}^0$ mixing, and perhaps actually observe it.
Therefore, it is time to take a detail look of all possible
techniques for $D^0\bar{D}^0$ mixing search.

This paper~\footnote{
This paper is essentially a revised version of Chapter 6 in~\cite{Thesis}
and ~\cite{Charm2000_liu}.}
is organized as follows: in Section 2 there is a review of the
experimental techniques which can be used to search for mixing,
together with some thoughts on possible new techniques.
In each case, the relevant phenomenology will be briefly presented.
Section 3 discusses the history, present status and future prospects of
searching for mixing at different experiments. In Section 4,
a comparison of the future prospects of the
different experiments with different
techniques, in the light of the CLEO II signal for $D^0\to K^+\pi^-$,
will be given. A brief summary is given in Section 5.
Some detailed formulae and discussions (including possible CP violation
effect) are provided in the appendices.

\begin{figure}[p]
\unitlength 1in
\begin{picture}(6.5,6)(0,0)
\put(-.45,-1.25){\psfig{width=6.96in,height=9.5in,%
file=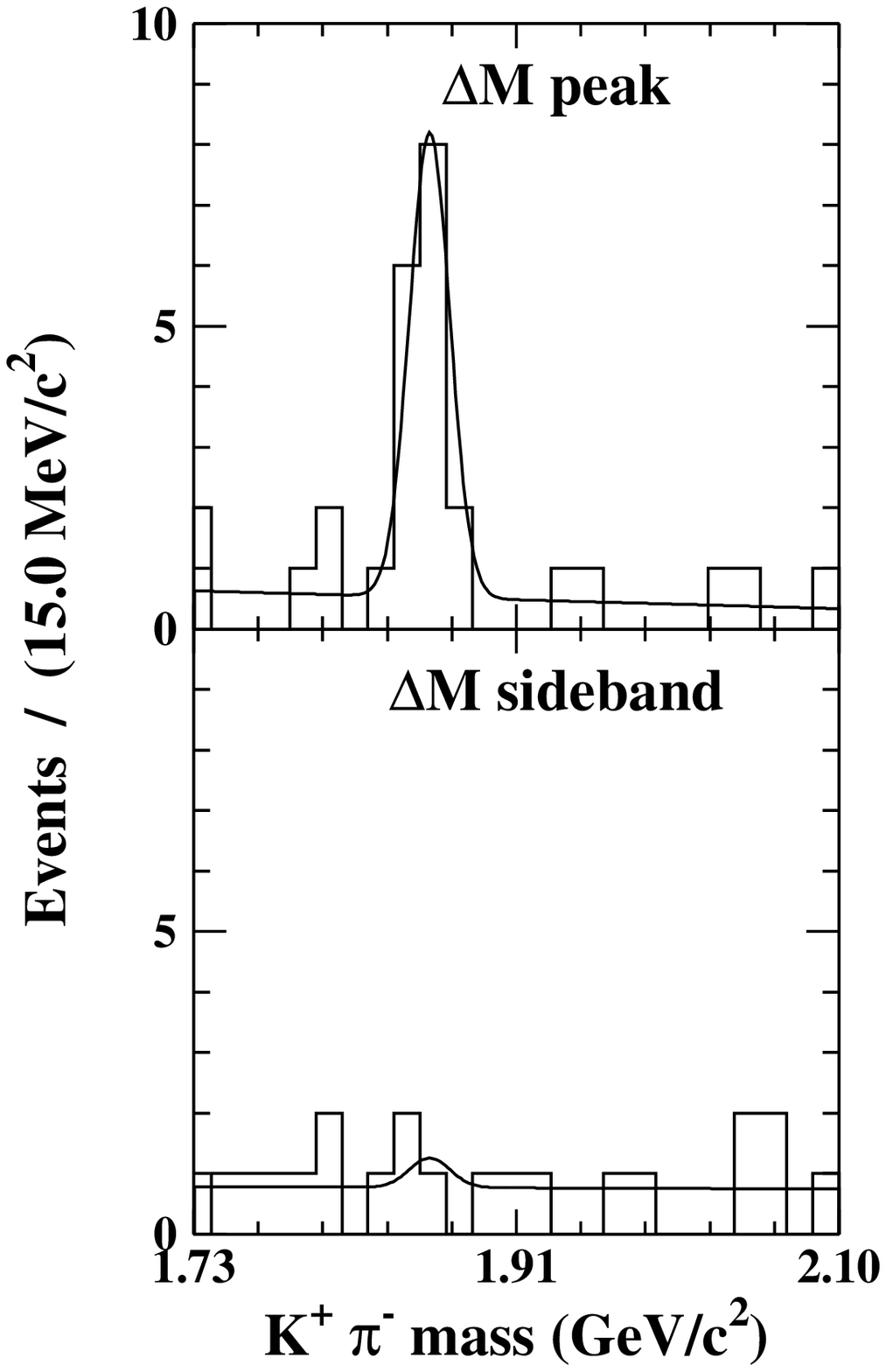}}
\end{picture}
\caption {The CLEO II signal for $D^0 \to K^+\pi^-$.
The $D^0$ mass for wrong sign events. (a) for events
in the $\Delta M$ peak; (b) for events in the
$\Delta M$ sidebands. The solid lines are the fits using the
corresponding right sign mean and $\sigma$ in data.}
\label{signal}
\end{figure}

\section{The Techniques}

The techniques which can be used to search for mixing can be roughly
divided into two classes: hadronic and semi-leptonic. Each method has
advantages and limitations, which are described below.

\subsection{Hadronic method}
The hadronic method is to search for the $D^0$ decays
$D^0\to K^+\pi^-(X)$. These decays can occur either through
$D^0\bar{D}^0$ mixing followed by Cabibbo favored
decay $D^0 \to \bar{D}^0  \to K^+\pi^-$(X), or
through DCSD
$D^0\to K^+\pi^-(X)$. This means that the major complication for this
method is the need to distinguish between DCSD and mixing~\cite{early}.
The hadronic method can therefore be classified according to how
DCSD and mixing are distinguished. In principle, there are
at least three different ways to distinguish between DCSD and
mixing candidates experimentally:
(A) use the difference in the decay time-dependence;
(B) use the possible difference in the resonant substructure
    between DCSD and mixing events
    in $D^0 \to K^+\pi^-\pi^0$,$ K^+\pi^-\pi^+\pi^-$, etc. modes;
(C) use the quantum statistics of the production
and the decay processes.

Method (A) requires that the $D^0$ be highly boosted and so that
the decay time information can be measured. Method (B) requires
knowledge of the resonant
substructure of the DCSD decays, which is unfortunately something
about which we have no idea at this time.
Finally, method (C) requires that one use $e^+e^-$
annihilation in the charm threshold region.
In the following, we will discuss these three methods in some detail.

\subsubsection{Method A --use the
difference in the time-dependence of the decay}

This method~\cite{Bjorken} is to measure the decay time of the
$D^0 \to K^+\pi^-$ decay.
Here the $D^0$ tagging is usually done by using
the decay chain $D^{*+} \to D^0{\pi}_{\rm s}^+$ followed by
$D^0 \to K^+ \pi^-$. The ${\pi}_{\rm s}^+$ from $D^{*+}$ has a soft momentum
spectrum and is referred to as ``the slow pion''.
The idea is to search for the wrong sign $D^{*+}$ decays, where the
slow pion has the same charge as the kaon arising from the $D^0$ decay.
This technique utilizes the following facts:
(1) DCSD and mixing have different decay time-dependence, which will
be described below.
(2) The charge of the
slow pion is correlated with the charm quantum number
of the $D^0$ meson and thus can be used to tag whether a $D^0$ or
$\bar{D}^0$ meson was produced in the decay
$D^{*+} \to D^0{\pi}_{\rm s}^+$ or $D^{*-} \to \bar{D}^0{\pi}_{\rm s}^-$.
(3) The small $Q$ value of the $D^{*+}$ decay
results in a very good mass resolution in the mass difference
$\Delta M \equiv M(D^{*+}) - M(D^0) - M({\pi}_{\rm s}^+)$ and allows
a $D^{*+}$ signal to obtained with very low background.
(4) The right sign signal $D^{*+} \to D^0{\pi}_{\rm s}^+$
followed by $D^0 \to K^- \pi^+$ can be used to provide
a model-independent normalization for the mixing measurement.

A pure $D^0$ state generated at $t=0$  decays to the $K^+\pi^-$ state either
by $D^0\bar{D}^0$ mixing or by DCSD, and the
two amplitudes may interfere.
The amplitude for a $D^0$
decays to $K^+\pi^-$ relative to the
amplitude for a $D^0$ decays to $K^-\pi^+$ is given by
(see appendix A)
\begin{equation}
\label{mixingamp}
A= \sqrt{{\rm R}_{\rm mixing}/2}\:\; t +
\sqrt{{\rm R}_{\rm DCSD}}\:\; e^{i\phi}
\end{equation}
where $\phi$ is an unknown phase, t is
measured in units of average $D^0$
lifetime. Detailed discusion on the interference phase $\phi$ can be found
in Appendix A.
Here ${\rm R}_{DCSD}=|\rho|^2$
where $\rho$ is defined as:
\begin{equation}
\label{rho_dcsd}
 \rho = {Amp(D^0\to K^+\pi^-) \over
           Amp(\bar{D}^0\to K^+\pi^-) }
\end{equation}
denoting the relative strength of DCSD.
We have also assumed a small mixing; namely,
$\delta m, \gamma_- \ll \gamma_+$ or $x,y\ll 1$, and
CP conservation. Detailed formulae and discussions (including
possible CP violation effect) can be found in Appendix A.
In the following, we will simply discuss the basic idea of
how to distinguish DCSD and mixing with this technique.

The first term, which is proportional to $t$, is due to mixing
and the second term is due to DCSD. It is this unique attribute of the
decay time-dependence of mixing which can be used to distinguish
between DCSD and mixing. Now we have:
\begin{equation}
\label{mixing}
{\rm I}(D^0 \to K^+\pi^-)(t) \propto ({\rm R}_{\rm DCSD}
+\sqrt{2{\rm R}_{\rm mixing}{\rm R}_{\rm DCSD}}\;\: t\: cos\phi
+\frac{1}{2}\;{\rm R}_{\rm mixing}t^2)e^{-t}
\end{equation}
Note that this form is different
from what people usually use (but equivalent), see Appendix A.
I prefer this form since
it is not only more convenient for discussion here, but also much
easier to be used to fit data.
Define $\alpha = {\rm R}_{\rm mixing}/{\rm R}_{\rm DCSD}$, which
describes the strength of mixing relative to DCSD. Equation~\ref{mixing}
can then be rewritten as:
\begin{equation}
\label{mixing1}
{\rm I}(D^0 \to K^+\pi^-)(t) \propto {\rm R}_{\rm DCSD}(1+\sqrt{2\alpha}
\; t cos\phi +\frac{1}{2}\alpha t^2)e^{-t}
\end{equation}

{}From this equation, one may read off the following
properties~\cite{Charm2000_liu}:
(1) The mixing term peaks at $t=2$.
(2) The interference term peaks at $t=1$.
(3) A small mixing signature can be enhanced by DCSD
through interference (with $cos\phi \neq 0$) at lower
decay times, compared to the case without interference (with $cos\phi=0$).
The ratio between the interference term and the mixing
term, denoted $\xi (t)$, is given by
$\xi (t)=\sqrt{\frac{8}{\alpha}}$ $cos\phi/t \propto \sqrt{\frac{1}{\alpha}}$.
So
when $\alpha \rightarrow 0$, $\xi \rightarrow \infty$.
(4) Only for $t>\sqrt{\frac{8}{\alpha}}|cos\phi|$ does the interference
term become smaller than the mixing term.
(5) ${\rm I}(t_{0})=0$ happens and only happens when $cos\phi=-1$,
and only at location $t_{0}=\sqrt{\frac{2}{\alpha}}$.
(6) One can obtain a very pure DCSD sample by cutting at low decay time.

While Property (1) tells us that the mixing term does live at longer
decay time,
Property (3) tells us clearly that we should not ignore the interference
term. In fact, that's the last thing one wants to ignore! (unless
we know for sure $\cos\phi=0$).
The commonly believed ``annoying background'', namely DCSD,
could actually enhance the chance of seeing a very small mixing signal
through the interference, compared to the case without the
interference. In other words, the ``annoying DCSD background''
does not necessary inherently limit the hadronic method since
the potentially small mixing signature could show up in the
interference term.
For a very small mixing rate, almost
all the mixing signature could show up in the interference term, not
in the mixing term, as long as $\cos\phi \neq 0$.
Property (2) tells us at which location one expect to find
the richest signature of a potential small mixing, which is
where the interference term peaks: $t \sim 1$
(why should one keep worrying about long lived DCSD tails?
let's hope for $\cos\phi \neq 0$ first).
Property (5) shows that destructive interference is not necessarily
a bad thing.
In fact, it could provide extra information. For example, if
$\cos\phi = -1$, then one should find ${\rm I}(t_{0})=0$
at $t_{0}=\sqrt{\frac{2}{\alpha}}$, see Figure~\ref{alpha}. Note this
unique attribute will become more interesting in method B, see Appendix B.
This tells us that the destructive interference does not necessarily
wipe out the signature of mixing.
For the general case, interference will lead to very characteristic time
distribution, as can be clearly seen in Figure~\ref{cosphi}.
Property (6) shows that we can study DCSD well
without being confused by the
possible mixing component. This will also become more important
when we discuss method B.

\begin{figure}[p]
\unitlength 1in
\begin{picture}(6.5,6)(0,0)
\put(-.45,-1.25){\psfig{width=6.96in,height=9.5in,%
file=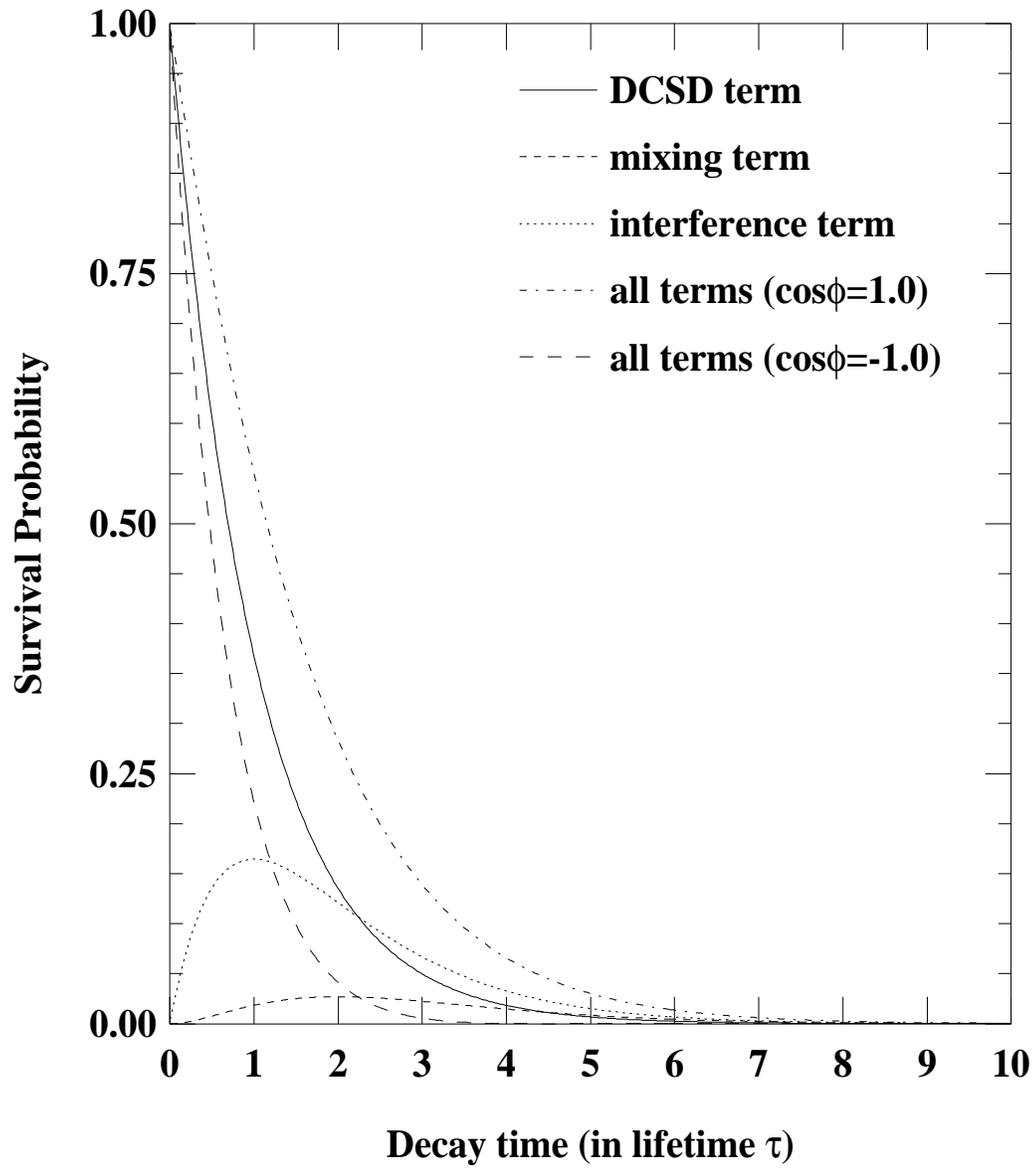}}
\end{picture}
\caption {The decay time dependence of DCSD and mixing with
$\alpha = {\rm R}_{\rm mixing}/{\rm R}_{\rm DCSD}=10\%$.}
\label{decaytime}
\end{figure}

\begin{figure}[p]
\unitlength 1in
\begin{picture}(6.5,6)(0,0)
\put(-.45,-1.25){\psfig{width=6.96in,height=9.5in,%
file=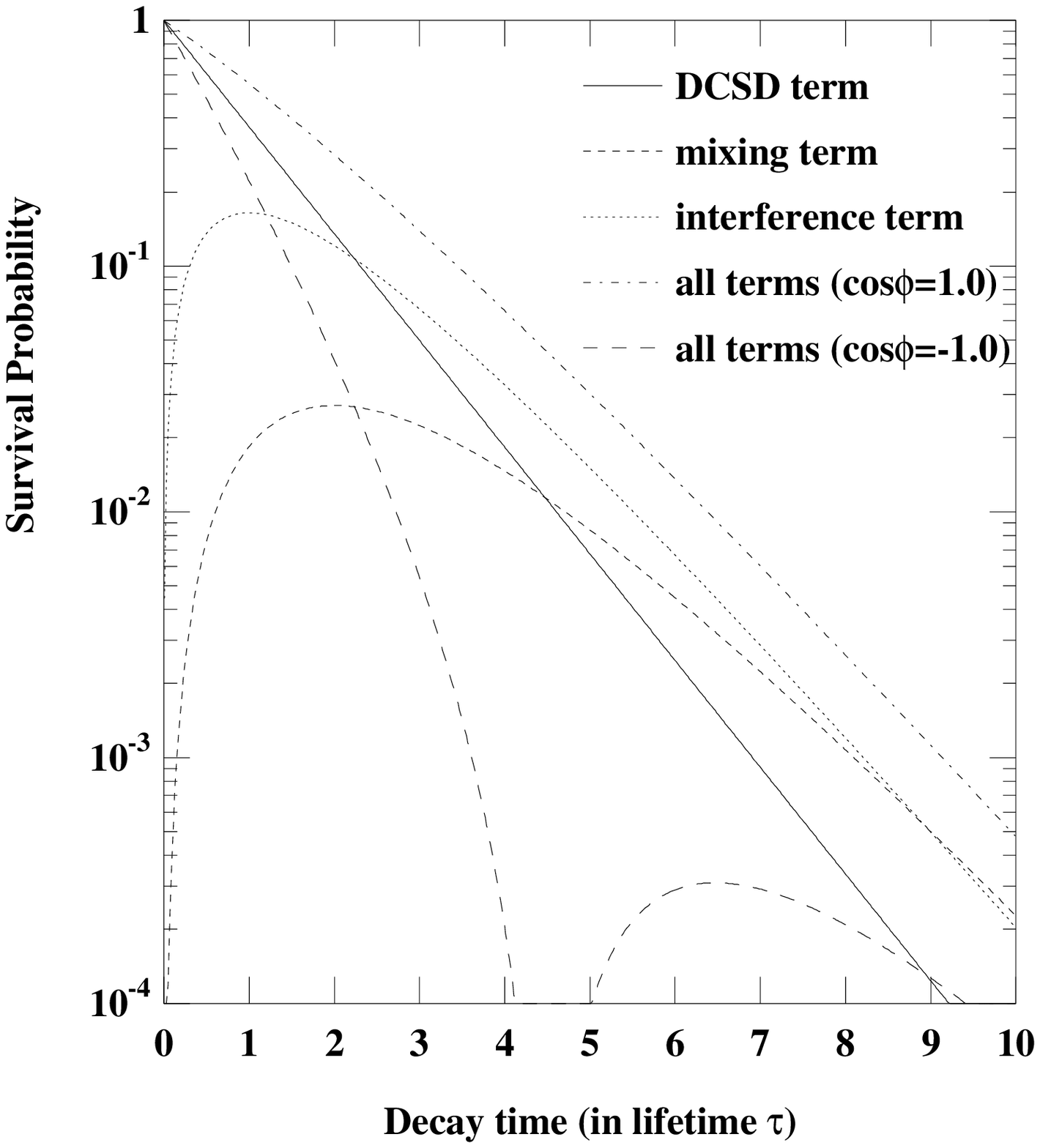}}
\end{picture}
\caption {The decay time dependence of DCSD and mixing
with $\alpha = {\rm R}_{\rm mixing}/{\rm R}_{\rm DCSD}=10\%$, in log scale.}
\label{decaytimelog}
\end{figure}

Therefore the signature of mixing is
a deviation from a perfect exponential time distribution with the
slope of $\gamma_+$~\footnote{One can use $D^0 \to K^-\pi^+$
to study the acceptance function versus decay time.}.
Our ability to observe this signature depends on
the number of $D^0 \to K^+\pi^-$ events we will have.
Right now this is limited by the rather poor statistics.
Figures~\ref{decaytime} or Figure~\ref{decaytimelog}
shows each term with $\alpha = 10\%$ and $cos\phi = \pm 1$
(with ${\rm R}_{\rm DCSD}=1$).

It is worth to point out that the interference between mixing and DCSD
also occurs in $B^0\bar{B^0}$ system.
In this case, mixing is quite large and can be well measured
while DCSD is small and unknown.
The signature of the small DCSD would
mostly show up in the interference term. But here we are not interested
in measuring mixing nor measuring DCSD, what is interesting here is to
measure CP violation. In Appendix C, We will use
$B^0_{d} \to D^+\pi^-$ as an example to show the basic idea.

It is interesting to point out here that there is also a possibility,
previously unrecognized, of using the Singly Cabibbo
Suppressed Decays (SCSD), such as $D^0 \to K^+K^-,\pi^+\pi^-$
to study mixing~\cite{Charm2000_liu}. This is because (assuming CP
conservation)
those decays occur only through the CP even eigenstate,
which means the decay time distribution is a perfect
exponential with the slope of $\gamma_1$.
Therefore, one can use those modes to measure $\gamma_1$.
The mixing signature is not a deviation from
a perfect exponential (again assuming CP conservation),
but rather a deviation of the
slope from $(\gamma_1 + \gamma_2)/2$. Since
${\gamma_+}= (\gamma_1 + \gamma_2)/2$ can be measured by
using the $D^0 \to K^-\pi^+$ decay time distribution, one can then
derive
$y={\gamma_- / \gamma_+}=(\gamma_2 - \gamma_1) /(\gamma_1 + \gamma_2)$.
Observation of a non-zero $y$ would demonstrate mixing caused
by the decay rate difference
(${\rm R}_{\rm mixing}= (x^2 + y^2)/2 $).
It is worth pointing out
that in this case other CP even (odd) final states
such as $D^0 \to K_{S} \rho^0$
can be also used to measure $\gamma_1 (\gamma_2)$.
In addition, there is no need to tag the $D^0$, since we only need to determine
the slope. Note that this method is only sensitive to
mixing caused by the decay rate difference between the two eigen states,
not to mixing caused by the mass difference $x={\delta m / \gamma_+}$
($\delta m = m_2 - m_1$).
Right after this technique was introduced~\cite{Charm2000_liu} last
summer, Fermilab fixed target experiments E791 started to
apply this idea to their data~\cite{Peng}.
The sensitivity of this method is discussed in Section 4.1.

\begin{figure}[p]
\unitlength 1in
\begin{picture}(6.5,6)(0,0)
\put(-.45,-1.25){\psfig{width=6.96in,height=9.5in,%
file=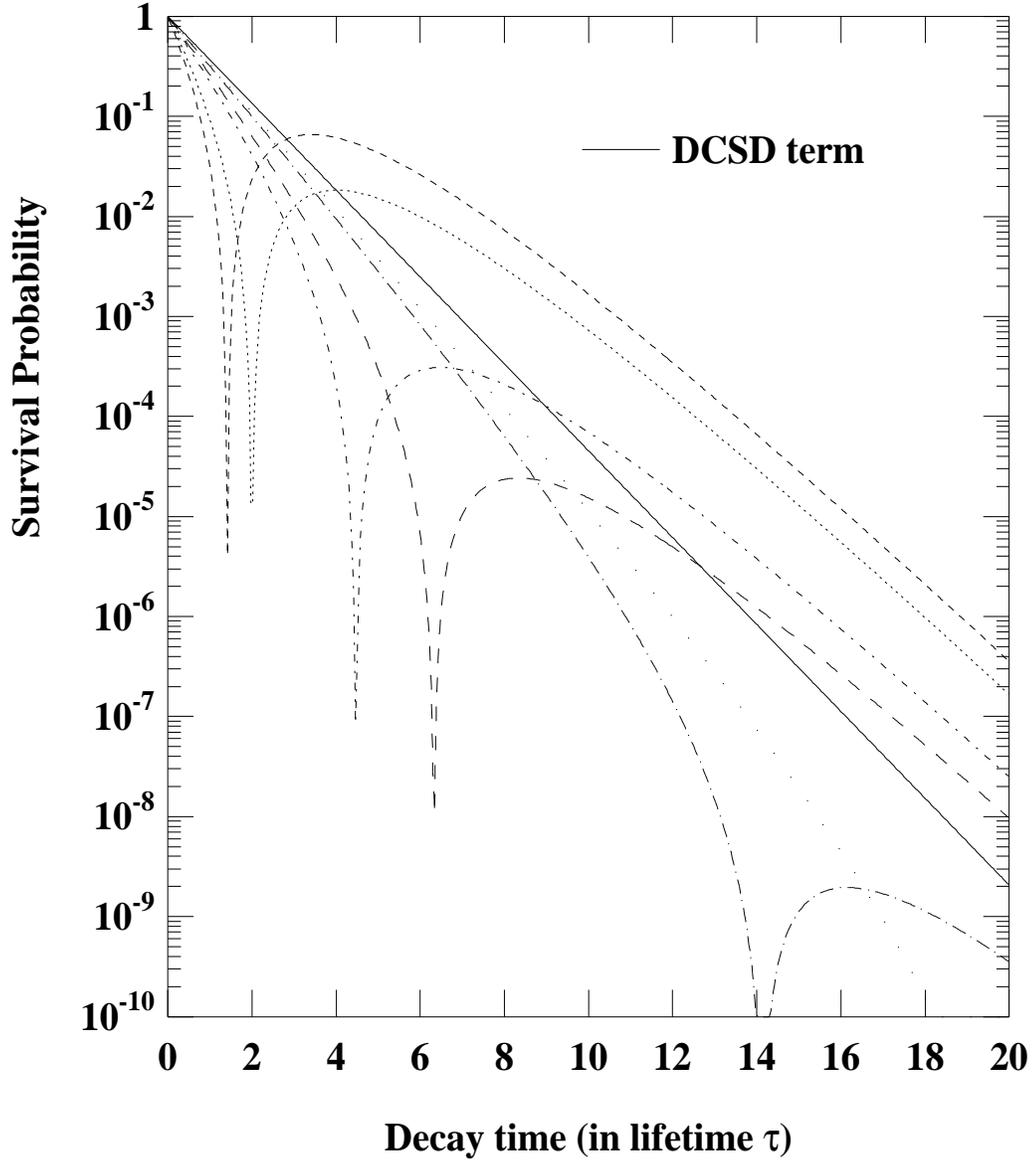}}
\end{picture}
\caption {The decay time dependence of DCSD and mixing
with maximal destructive interference $cos\phi=-1.0$.
For different $\alpha = {\rm R}_{\rm mixing}/{\rm R}_{\rm DCSD}$ values:
from left to right, $\alpha=100\%, 50\%, 10\%$,
$5\%, 1\%,0.5\%$ (with ${\rm R}_{\rm DCSD}=10^{-2}$,
this corresponds to ${\rm R}_{\rm mixing}=$ $10^{-2},
5 \times 10^{-3},10^{-3}$,
$5 \times 10^{-4},10^{-5},5 \times 10^{-6}$).}
\label{alpha}
\end{figure}

\begin{figure}[p]
\unitlength 1in
\begin{picture}(6.5,6)(0,0)
\put(-.45,-1.25){\psfig{width=6.96in,height=9.5in,%
file=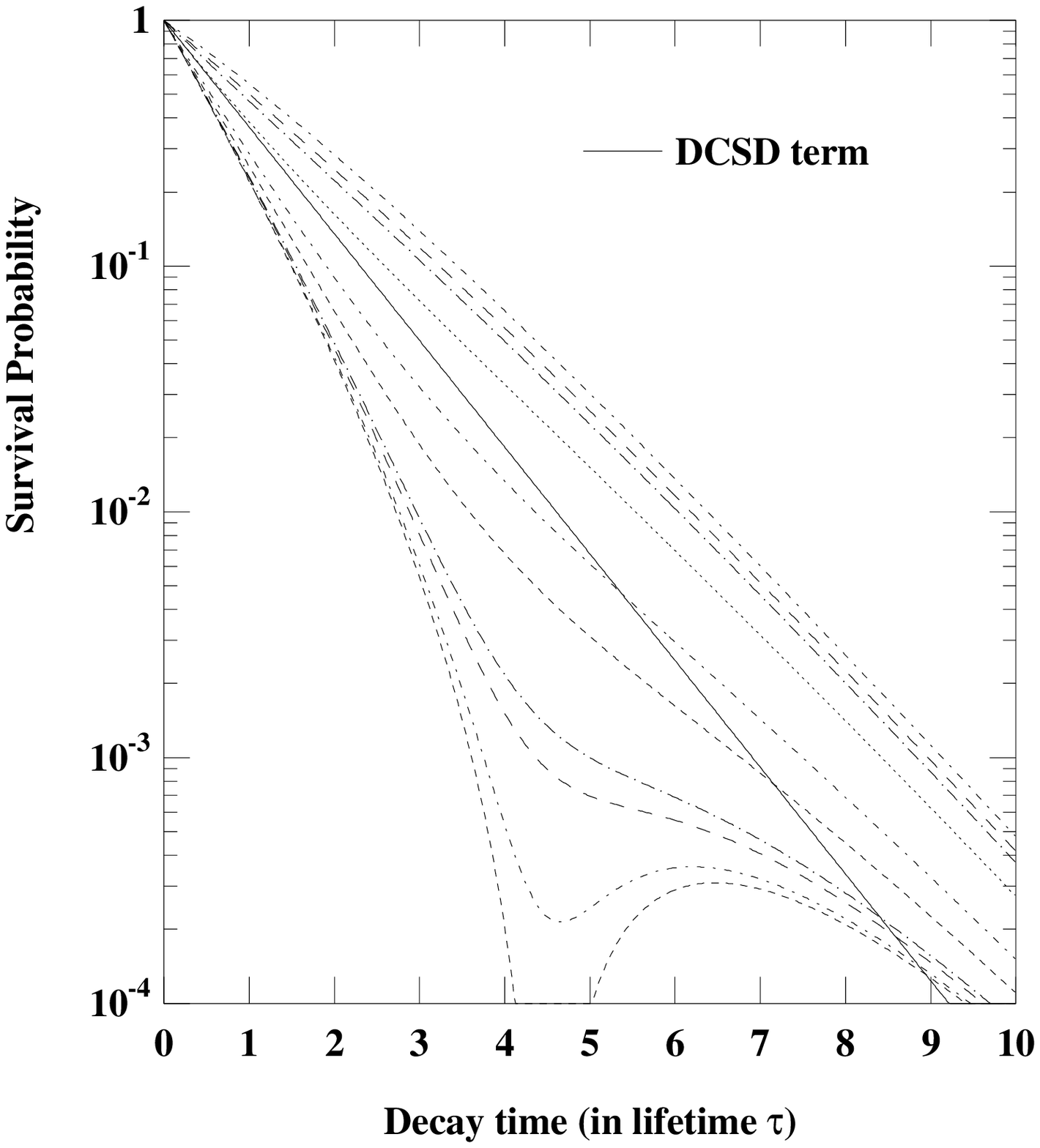}}
\end{picture}
\caption {The decay time dependence of DCSD and mixing
with $\alpha = {\rm R}_{\rm mixing}/{\rm R}_{\rm DCSD}=10\%$.
For different $cos\phi$ values: from bottom to top,
$cos\phi$ $=-1.0,-0.99,-0.96,-0.94,-0.80,$
$-0.6, 0.0, 0.5, 0.7, 1.0$. The solid line is the DCSD term,
as a reference line.}
\label{cosphi}
\end{figure}

\subsubsection{Method B -- use difference in resonance substructure}

The idea of this new method~\cite{Charm2000_liu} is to use the wrong sign
decay $D^{*+} \to D^0{\pi}_{\rm s}^+$ followed by
$D^0 \to K^+ \pi^-\pi^0$, $K^+\pi^-\pi^+\pi^-$, etc., and use the
possible differences of the resonant substructure
between mixing and DCSD to study mixing.
There are good reasons to believe that the resonant substructure
of DCSD decay is different from that of mixing (Cabibbo favored
decay, CFD).
We can use the $D^0 \to K^+ \pi^-\pi^0$ decay as an example.
Detail discussion about this method can be found in appendix B (including
possible CP violation effect),
here we will just outline the basic idea.

For CFD and DCSD,
the true yield density $n(p)$ at a point $p$ in the Dalitz plot
can be written as:
\begin{equation}
\label{density}
n(p) \:\propto \: \vert f_{1}\:e^{i\phi_{1}}A_{3b} +
f_{2}\:e^{i\phi_{2}}BW_{\rho^{+}}(p) + f_{3}\:e^{i\phi_{3}}BW_{K^{*-}}(p)
+f_{4}\:e^{i\phi_{4}}BW_{\bar{K}^{*0}}(p) {\vert}^2
\end{equation}
where $f_{i}$ are the relative amplitudes for each component and $\phi_{i}$
are the interference phases between each submode.
$A_{3b}$ is the S-wave three-body
decay amplitude, which is assumed to be
flat across the Dalitz plot. The various
terms $BW$ are Breit-Wigner amplitudes for the $D^0 \to K^{*}\pi$ and
$D^0 \to K\rho$ sub-reactions,
which describe the strong resonances and decay angular momentum conservation:
$BW_{R} \:\propto \: \frac{\cos\theta_{R}}{M_{ij}-M_{R}-i\Gamma_{R}/2}$ where
$M_{R}$ and $\Gamma_{R}$ are the mass and width of the $M_{ij}$
resonance ($K^{*}$ or $\rho$), and $\theta_{R}$ is the helicity angle of
the resonance.
For CFD, $f_{i}$ and $\phi_{i}$
have been measured by MARKIII~\cite{Weinstein}, E691~\cite{E6911}
and are being measured by CLEO II. For DCSD, $f_{i}$ and $\phi_{i}$
have not been measured.
Note that in general
\begin{equation}
\label{dcsd}
{f_{i}}^{DCSD}/{f_{i}}^{CFD}\neq {f_{j}}^{DCSD}/{f_{j}}^{CFD} \; (i \neq j)
\end{equation}
\begin{equation}
\label{phase}
{\phi_{i}}^{DCSD} \neq {\phi_{i}}^{CFD}
\end{equation}
This means that the resonant
substructure (the true yield density $n(p)$) for DCSD is different
from that of
mixing.
As both DCSD and mixing contribute to the wrong sign
decay, the yield density for the wrong sign events
$n_w(p)$ will have a complicated form.
Just like in method A, for very small mixing, the interference
term between DCSD and mixing could be the most important one.

Mathematically, the time-dependence of $D^0 \to K^+\pi^-\pi^0$
is the same as that of
$D^0 \to K^+\pi^-$, the only difference is that now both
${\rm R}_{\rm DCSD}$ and the interference phase $\phi$ (between
DCSD and mixing) strongly depends on the location $p$ on the Dalitz plot.
As discussed in Appendix B, the time-dependence can
be written in the form~\cite{Thesis}:
\begin{eqnarray}
\label{kpipi0}
\lefteqn{I(~|D^0_{\rm phys}(t)> \to f~)~(p)=}  \nonumber \\
  & & {\left[~n_{\rm D}(p)
+ \sqrt{2{\rm R}_{\rm mixing}~n_{\rm D}(p)~n_{\rm C}(p)}~\cos\phi(p)~t
+ \frac{1}{2}~n_{\rm C}(p)~{\rm R}_{\rm mixing}~t^2\right]~e^{-t}. }
\end{eqnarray}
where $n_{\rm D}(p)$ and $n_{\rm C}(p)$ are the true yield density
for DCSD and CFD respectively. Detailed discussion on the interference
phase $\phi$ can be found in Appendix B.

In principle, one can use the difference between the
resonant substructure for DCSD and mixing events to
distinguish mixing from DCSD. For instance,
combined with method A, one can perform a multi-dimensional
fit to the data by using the information on
$\Delta M$, $M(D^0)$, proper decay time $t$ and
the yield density on Dalitz plot
$n_{w}(p,t)$. The extra information on the
resonant substructure will, in principle, put a much better
constraint on the amount of mixing. Of course,
precise knowledge of the resonant substructure for DCSD
is needed here and so far we do not know anything
about it. Because of this, for current experiments
this method is more likely to be a complication
rather than a better method
when one tries to apply method A to
$D^0 \to K^+ \pi^-\pi^0$ (see~\cite{Liudpf} and~\cite{Thesis})
or $D^0 \to K^+ \pi^-\pi^+\pi^-$.
In principle, however, one can use wrong sign samples at
low decay time (which is almost pure DCSD) to
study the resonant substructure of the DCSD decays.

It is interesting to point out here (as discussed in detail
in Appendix B) that the Dalitz plot changes its shape as the decay time
``goes by'' due to the interference effect. Note that the
interference phase, unlike in the case of $D^0 \to K^+\pi^-$,
strongly depends on the location on the Dalitz plot since
there are contributions from the various Breit-Wigner
amplitudes, which changes wildly across each resonance.
One would expect that $\cos\phi(p)$ could have any value
between [-1,1], depending on the location $p$. It is interesting
to look at the locations on the Dalitz plot where
$\cos\phi(p)=-1$ (maximal destructive mixing-DCSD interference).
As pointed out in method A, Property (5) tells us that
one should find ${\rm I}(t_{0})=0$ at $t_{0}=\sqrt{\frac{2}{\alpha}}$.
This means that the mixing-DCSD interference would dig a ``hole''
on the Dalitz plot at time $t_{0}$ at that location. Since
$t_0 \propto \sqrt{{\rm R}_{\rm DCSD}(p)}=\sqrt{n_{\rm D}(p)/n_{\rm C}(p)}$,
the ``holes'' would show up earlier (in decay time) at locations where
${\rm R}_{\rm DCSD}(p)$ is smaller. Imagine that someone watches the
Dalitz plot as the decay time ``goes by'', this person
would expect to see ``holes'' moving
from locations with $\cos\phi(p)=-1$ and smaller
${\rm R}_{\rm DCSD}(p)$ toward locations with
$\cos\phi(p)=-1$ and larger ${\rm R}_{\rm DCSD}(p)$.
The existence of the ``moving holes'' on the Dalitz plot would
be clear evidence for mixing. Once again we see the importance
of the mixing-DCSD interference effect.

In the near future, we should have a good understanding
of DCSD decays and this method could become a feasible
way to search for mixing (and CP violation).

\subsubsection{Method C ---use quantum statistics of the production
and decay processes}
This method is to search for dual identical two-body hadronic
decays in $e^+e^- \to \Psi'' \to D^0\bar{D}^0$, such as
$(K^-\pi^+)(K^-\pi^+)$, as was first suggested by Yamamoto
in his Ph.D thesis~\cite{Yamamoto}.
The idea is that when $D^0\bar{D}^0$ pairs are generated in a state
of odd orbital angular momentum (such as $\Psi''$), the DCSD
contribution to identical two-body pseudo-scalar-vector ($D \to PV$)
and pseudo-scalar-pseudo-scalar ($D \to PP$) hadronic decays
(such as $(K^-\pi^+)(K^-\pi^+)$) cancels out, leaving only the contribution
of mixing~\cite{Yamamoto,Bigi,Du}.
The essence of Yamamoto's
original calculation for the
$(K^-\pi^+)(K^-\pi^+)$ case is given below.

Let's define $e_i(t)=e^{-im_it- \gamma_i t/2}$ ($i=1,2$) and
$e_{\pm}(t) = (e_1(t) \pm e_2(t))/2$. A state that is
purely $|D^0 \rangle $ or $| \bar{D}^0 \rangle$ at time $t=0$ will evolve to
$|D(t) \rangle$ or $| \bar{D}(t)\rangle$ at time $t$, with
$|D(t) \rangle=e_+(t) |D^0 \rangle + e_-(t) |\bar{D}^0 \rangle$
and $|\bar{D}(t) \rangle = e_-(t) |D^0\rangle + e_+(t) |\bar{D}^0\rangle$.
In $e^+e^- \to \Psi'' \to D^0\bar{D}^0$,
the $D^0\bar{D}^0$ pair is generated in the state
$D^0\bar{D}^0 - \bar{D}^0D^0$ as the
relative orbital angular momentum of the pair
${\cal L} = 1$. Therefore, the time evolution of this state
is given by $|D(t)\bar{D}(t')\rangle - | \bar{D}(t)D(t') \rangle$,
where $t$ ($t'$) is the time of decay of the $D$ ($\bar{D}$).
Now the double-time amplitude ${\cal A}_w (t,t')$
that the left side decays to $K^-\pi^+$ at $t$ and
the right side decays to $K^-\pi^+$ at $t'$, giving
a wrong sign event $(K^-\pi^+)(K^-\pi^+)$, is given by:
\begin{equation}
\label{ddbarampw}
{\cal A}_w(t,t') = (e_+(t)e_-(t') - e_-(t)e_+(t'))(a^2 - b^2)
\end{equation}
where $a=\langle K^-\pi^+ | D^0 \rangle$ is the amplitude of the Cabibbo
favored decay $D^0 \to K^-\pi^+$, while $b=\langle K^-\pi^+|\bar{D}^0 \rangle$
is the amplitude of DCSD
$\bar{D}^0 \to K^-\pi^+$. Similarly, the double-time
amplitude ${\cal A}_r(t,t')$ for the right sign event
$(K^-\pi^+)(K^+\pi^-)$ is given by:
\begin{equation}
\label{ddbarampr}
{\cal A}_r(t,t') = (e_+(t)e_+(t') - e_-(t)e_-(t'))(a^2 - b^2)
\end{equation}
One measures the wrong sign versus right sign ratio ${\rm R}$, which is:
\begin{equation}
\label{nodcsd}
{\rm R}=\frac{{\rm N}(K^-\pi^+,K^-\pi^+)+{\rm N}(K^+\pi^-,K^+\pi^-)
}{{\rm N}(K^-\pi^+,K^+\pi^-)+{\rm N}(K^+\pi^-,K^-\pi^+) }
=\frac{\int\!\!\int |{\cal A}_w(t,t')|^2\,dt\,dt'}
{\int\!\!\int |{\cal A}_r(t,t')|^2\,dt\,dt'}
\end{equation}
Note in taking the ratio, the amplitude term $(a^2 - b^2)$ in
Equations~\ref{ddbarampw}
and~\ref{ddbarampr} drops out. Thus, clearly $\rm R$ does not depend on
whether $b$ is zero (no DCSD) or finite (with DCSD).
Integrating over all times, one then obtains
${\rm R}=(x^2 + y^2)/2 = {\rm R}_{\rm mixing}$, where $x$ and $y$
are defined as before.

This is probably
the best way to separate DCSD and mixing.
The exclusive nature of the production guarantees both
low combinatoric backgrounds and production kinematics
essential for background rejection.
This method requires one use $e^+e^-$ annihilation
in the charm threshold region. Here the best final state is
$(K^-\pi^+)(K^-\pi^+)$.
In principle, one can also use final states like
$(K^-\rho^+)(K^-\rho^+)$ or $(K^{*-}\pi^+)(K^{*-}\pi^+)$,
etc., although again there are complications. For example, it is
hard to differentiate experimentally $(K^-\rho^+)(K^-\rho^+)$ from
$(K^-\rho^+)(K^-\pi^+\pi^0)$, where DCSD can contribute.
With high statistics, in principle, this method could
be combined with method B.

It has been pointed out that quantum statistics yield
different correlations for the $D^0\bar{D}^0$ decays from
$e^+e^- \to D^0\bar{D}^0, D^0\bar{D}^0\gamma,D^0\bar{D}^0\pi^0$~\cite{Bigi2}.
The well-defined coherent quantum states of the $D^0\bar{D}^0$
can be, in principle, used to provide valuable cross checks on systematic
uncertainties, and to
extract $ x={\delta m / \gamma_+}$ and $ y={\gamma_- / \gamma_+}$
(which requires running at different energies)
if mixing is observed~\cite{Bigi2}.

\subsection{Semi-leptonic method}
The semi-leptonic method is to search for
$D^0\to \bar{D^0}\to X l^- \nu$ decays,
where there is no DCSD involved. However, it usually (not always!)
suffers from a large background due to the missing neutrino.
In addition, the need to
understand the large background often introduces model dependence.
In the early days, the small size of fully
reconstructed samples of exclusive $D^0$ hadronic decays and the lack of
the decay time information made it difficult to constrain the
$D^0\bar{D}^0$ mixing rate using the hadronic
method, many experiments used semi-leptonic decays.
The techniques that were used were similar
---searching for like-sign
$\mu^+\mu^+$ or $\mu^-\mu^-$ pairs in
$\mu^+N\to \mu^+(\mu^+\mu^+)X$~\cite{EMC,BDMS} and $\pi^- Fe \to
\mu^+\mu^+$~\cite{CCFRS}, $\pi^- W \to \mu^+\mu^+$~\cite{E615}.
These techniques rely
on the assumptions on production mechanisms, and the accuracy
of Monte Carlo simulations to determine the large conventional sources
of background.

There are other ways of using the semi-leptonic method.
The best place to use the semi-leptonic method is probably in
$e^+e^-$ annihilation near the charm threshold region.
The idea is to search for $e^+e^- \to \Psi'' \to $
$D^0\bar{D}^0 \to (K^-l^+\nu)(K^-l^+\nu)$ or
$e^+e^- \to D^-D^{*+} \to (K^+\pi^-\pi^-)(K^+\l^-\nu){\pi}_{\rm s}^+$
{}~\cite{Gladding3,Schindler}.
The latter is probably the only place where the semi-leptonic method does
not suffer from a large background. It should have a low background,
as there is only one neutrino missing in the entire event,
threshold kinematics constraints should provide clean signal.

It has been pointed out that one can not claim a
$D^0\bar{D}^0$ mixing signal based on the semi-leptonic method
alone (unless with the information on decay time of $D^0$).
Bigi~\cite{Bigi2} has pointed out that an observation of
a signal on $D^0 \to l^-X$ establishes only that a certain selection
rule is violated in processes where the charm quantum number is changed,
namely the rule $\Delta {\rm {Charm}} = - \Delta {\rm Q_l}$ where
${\rm Q_l}$ denotes leptonic charge. This violation can occur
either through $D^0\bar{D}^0$ mixing (with the unique attribute of
the decay time-dependence of mixing), or through new physics beyond
the Standard Model (which could be independent of time).
Nevertheless, one can always use this method to set upper limit
for mixing.

\section{Mixing Searches at Different Experiments}

\subsection{$e^+e^-$ running on $\Psi''(3770)$ --MARK III, BES,
Tau-charm factory}
\label{review-exps-psi}

The MARK III collaboration was the first (though
hopefully not the last)
to use the $e^+e^- \to \Psi'' \to D^0\bar{D}^0$ technique.
They reported three events consistent with $|\Delta S = 2|$
transitions~\cite{Gladding1}.
One event is observed in the final state $K^+ \pi^-$ vs $K^+ \pi^-\pi^0$.
The other two are reconstructed in the final states
$K^+ \pi^-\pi^0$ vs $K^+ \pi^-\pi^0$, and a Dalitz plot analysis
finds one to be consistent with  $K^- \rho^+$ versus  $K^- \rho^+$
and the other consistent with $K^{*0}\pi^0$ versus $K^{*0}\pi^0$ (note the
$D^0 \to K^- \pi^+\pi^0$ decays are dominated by
$D^0 \to K^- \rho^+$ and $D^0 \to \bar{K}^{*0}\pi^0$ channels).
Using a maximum likelihood
analysis, they interpreted the results for two limiting cases: a).
if there is no DCSD in $D^0 \to K^+ \pi^-$ and $D^0 \to K^+ \pi^-\pi^0$;
then the events imply ${\rm R}_{\rm mixing}$ = $( 1.2\pm 0.6\, ) \% $ or
${\rm R}_{\rm mixing} > 0.4\%$ at $90\%$ C.L.;
b). if there is no DCSD for $D^0 \to K^+ \pi^-$ and
no $D^0 \bar{D}^0$ mixing, and also at least one of the  $K^+ \pi^-\pi^0$'s
in each of those two events are non-resonant, then
the results imply ${\rm R}_{\rm DCSD}$ =
$\Gamma(D^0 \to K^+ \pi^-\pi^0)/\Gamma(D^0 \to K^- \pi^+\pi^0) =
( 7\pm 4\, )\:\tan^4\theta_C $ or ${\rm R}_{\rm DCSD} > 1.9 \:\tan^4\theta_C$
at 90\% C.L..
This was a interesting result at that time, and had a strong influence
on the subject. However, one cannot draw a firm conclusion
about the existence of $D^0\bar{D}^0$ mixing based on these events.
There are at least two reasons:
(1) The background study has to rely on Monte Carlo
simulation of the PID (particle identification --
Time-of-Flight)~\footnote{In principle, one can use kinematics to
check whether the events are due to doubly misidentified
$D^0 \to K^-\pi^+\pi^0$: if one inverts the
$K^+$, $\pi^-$ assignments and recalculates the $D^0$ mass
(let's call this ${\rm M}_{\rm flip}$),
the ${\rm M}_{\rm flip}$ will not be within $D^0$
mass peak if it is real, unless the $K$ and $\pi$ momentum is close.
Unfortunately, ${\rm M}_{\rm flip}$
for all the three events are within $D^0$
mass peak. That's why one needs to totally rely on
PID.}. As Gladding has pointed out: ``These results must be
considered preliminary
because the calculation of the confidence level is sensitive to the
tails of PID distribution for the background''~\cite{Gladding2};
(2) Assuming that the Monte Carlo background study is correct, and that
the events are real, one still cannot claim the two events are due to mixing,
for example, the non-resonant decays $D^0 \to K\pi\pi^0$ may contribute to one
side of the pair in each of the events, in which DCSD can contribute.

The MARK III puzzle can be solved at a $\tau$-charm factory,
which is a high luminosity ($10^{33} cm^{-2}s^{-1}$) $e^+e^-$ storage ring
operating at center-of-mass energies in the range 3-5 GeV.
The perspectives for a $D^0\bar{D}^0$ mixing search at a
$\tau$-charm factory have been studied in some
detail~\cite{Gladding3,Schindler}.
I will outline here the most important
parts. The best way to search for mixing at $\tau$-charm factory
is probably to use
$e^+e^- \to \Psi'' \to D^0\bar{D}^0 \to (K^-\pi^+)(K^-\pi^+)$.
The sensitivity is not hard to estimate. Assuming a one year run
with a luminosity of $10^{33} cm^{-2}s^{-1}$, $5.8 \times 10^7$ $D^0$s
would be produced from $\Psi''$. Therefore about $9 \times 10^4$
$(K^-\pi^+)(K^+\pi^-)$ events would be produced.
About $40\%$ of them ($3.6 \times 10^4$)
could be fully reconstructed.
A study~\cite{Gladding3} has shown that the potential dominant background
comes from doubly misidentified $(K^-\pi^+)(K^+\pi^-)$, and if
TOF resolution is 120 ps, this background could
be kept to the level of one event or less. This means one could,
in principle, set an
upper limit at the $10^{-4}$ level.

As mentioned Section 2.2, the best place to use the semi-leptonic method
is probably at a $\tau$-charm factory.
One good example is to search for $e^+e^- \to D^-D^{*+} \to$
$(K^+\pi^-\pi^+)(K^+l^-\nu){\pi}_{\rm s}^+$. It is expected that
this method can also have a sensitivity at the $10^{-4}$ level.
There are many other independent techniques that one can use for a mixing
search at a $\tau$-charm factory. By combining several independent
techniques (which require running at different energies), it was claimed that
$D^0\bar{D}^0$ mixing at the $10^{-5}$ level could be
observable~\cite{Schindler}.

There have been several schemes around the world for building a
$\tau$-charm factory.
If such a machine is built, it could be
a good place to study mixing. The history of the
$\tau$-charm factory can be found in
reference ~\cite{Toki}: one was proposed at SLAC
in 1989 and one at Spain in 1993; one was discussed at
Dubna in 1991, at IHEP (China), and at Argonne~\cite{Argonne} in 1994 and
at this workshop.
It will be discussed again at IHEP (China) soon.
Let us hope that we will have one in the not-too-distant future.

\subsection{$e^+e^-$ running near $\Upsilon(4S)$ --ARGUS, CLEOII, CLEO III,
B factory}
\label{review-exps-upsilon}

Without a precision vertex detector, CLEO II can only in effect
measure the rate ${\cal B}(D^0\to K\pi)$
integrated over all times of a pure $D^0$ decaying to a final
state $K\pi$. The ratio
${\rm R}$=${\cal B}(D^0\to K^+ \pi^-)$ /${\cal B}(D^0 \to K^- \pi^+)$
is given by integrating equation~\ref{mixing} over all times
(see appendix A)
\begin{equation}
\label{labeled-equation}
{\rm R}= {\rm R}_{\rm mixing} + {\rm R}_{\rm DCSD}
+\sqrt{2{\rm R}_{\rm mixing}{\rm R}_{\rm DCSD}}\; cos\phi.
\end{equation}

CLEO finds~\cite{Liu} $ {\rm R}=
(0.77\pm 0.25\,({\rm stat.})\pm 0.25\,({\rm sys.}))\% $.
This signal could mean one of two things: (1) mixing could be quite large,
which would imply that mixing can be observed in the near future;
(2) the signal is dominated by DCSD.
The theoretical prediction for ${\rm R_{DCSD}}$ is about
$(2-3) tan^4\theta_C \sim (0.6-0.9)\%$~\cite{Bigi,Chau,Buccella},
which is quite
consistent with the measured value. It is, therefore,
believed by many that the signal is due to DCSD,
although it remains consistent with
the current best experimental upper limits on mixing, which
are $(0.37-0.7)\%$~\cite{Browder} and $0.56\%$~\cite{E615}.

CLEO has also tried to use hadronic method B, by
searching for $D^0 \to K^+\pi^-\pi^0$.
The excellent photon detection at CLEO II allows one to study this mode with
a sensitivity close to $D^0 \to K^+\pi^-$ mode.
The main complication faced here is that (as discussed in method B)
the resonant substructure is not
necessarily the same for wrong sign and right sign decays.
Because of this, as discussed in appendix B,
the interpretation of ${\rm R}$ as ${\rm R_{mixing}}$
or ${\rm R_{DCSD}}$ will be complicated by the lack of knowledge
of the details of the interference between submodes (and also the
decay time information).
Moreover, one has to worry about the detection efficiency
across the Dalitz plot. Setting an upper limit for each
submode is clearly very difficult.
CLEO has set an upper limit~\cite{Liudpf,Thesis} on the inclusive rate
for $D^0 \to K^+\pi^-\pi^0$ as
${\rm R}$ = ${\cal B}(D^0 \to K^+\pi^-\pi^0)$
/${\cal B}(D^0 \to K^-\pi^+\pi^0)$ $< 0.68\%$.
Note this upper limit includes the possible effects of the
interference between the DCSD and mixing for each submode
as well as the interference between submodes.

This summer, CLEO will install a silicon vertex detector (SVX)
with a longitudinal resolution on
vertex separation around 75 $\mu {\rm m}$. This will
enable CLEO to measure the
decay time of the $D^0$, and reduce the random slow pion background
(the resolution of the $D^{*+}$ - $D^0$ mass difference is dominated
by the angular resolution on the slow pion, which should be
greatly improved by the use of the SVX).
By the year 2000, with CLEO III (a symmetric B factory)
and asymmetric B factories at SLAC and KEK,
each should have thousands of $D^0 \to K^+K^-(X),\pi^+\pi^-(X)$
and a few hundred $D^0 \to K^+\pi^-$
(and perhaps $D^0 \to K^+\pi^-\pi^0$, $K^+\pi^-\pi^+\pi^-$ too)
signal events with decay time information for one year of running.
The typical decay length of
$D^0$ (${\cal L}$) is about a few hundred $\mu {\rm m}$, and the resolution
of the decay length ($\sigma_{\cal L}$) is about 80 $\mu {\rm m}$
(${\cal L}/{\sigma_{\cal L}} \sim 3$). The sensitivity to mixing at
CLEO III and asymmetric B factories has not been carefully studied yet.
A reasonable guess is that it could be as low as $10^{-4}$.
If mixing rate is indeed as large as DCSD rate,
it should be observed by then.

\subsection{Fixed target experiments}
\label{review-exps-fixed}

A significant amount of our knowledge has been gained from
Fermilab fixed target experiments, and in fact the current best upper
limits on mixing have emerged from these experiments (E615, E691), and
will come from their successors E687, E791 and E831 soon.

The best upper limit using the semi-leptonic method comes
from the Fermilab experiment E615, which used a 255 GeV pion beam on
a tungsten target. The technique is to search
for the reaction $\pi N \to D^0\bar{D}^0 \to$ $ (K^-\mu^+\nu)D^0$
$\to (K^-\mu^+\nu)(K^-\mu^+\nu)$, where only the final state
muons are detected (i.e. the signature is like-sign
$\mu^+\mu^+$ or $\mu^-\mu^-$ pairs).
Assuming $\sigma(c\bar{c}) \sim A^1$ nuclear dependence, they
obtained ${\rm R}_{\rm mixing} < 0.56\%$~\cite{E615}.

The best upper limit using the hadronic method by measuring the
decay time information comes from E691, which is the first high statistics
fixed target (photoproduction) experiment.
In fact, E691 was the first experiment which used the
decay time information (obtained from the excellent decay time
resolution of their silicon detectors) to distinguish DCSD and mixing.
The decay chains $D^{*+} \to D^0{\pi}_{\rm s}^+$ followed by
$D^0 \to K^+ \pi^-$, $K^+ \pi^-\pi^+\pi^-$ were used.
Their upper limits from the $D^0 \to K^+\pi^-$ mode
are ${\rm R}_{\rm mixing} < (0.5-0.9)\%$ and ${\rm R}_{DCSD} < (1.5-4.9)\%$ ,
while the upper limits from $D^0 \to K^+\pi^-\pi^+\pi^-$
are ${\rm R}_{\rm mixing}<(0.4-0.7)\%$ and ${\rm R}_{DCSD} < (1.8-3.3)\%$ .
The ranges above reflect the possible
effects of interference between DCSD and mixing with an unknown phase ($\phi$).
Although the combined result gives ${\rm R}_{\rm mixing} < (0.37-0.7)\%$,
in principle, one cannot combine the results from the two modes since the
interference phases are totally different, as discussed in appendix B.
Note that in their analysis for $D^0 \to K^+ \pi^-\pi^+\pi^-$,
the resonant substructure in the Cabibbo favored and
DCSD decays was ignored. As discussed in appendix B, in general, one
cannot treat
$D^0 \to K^+\pi^-\pi^+\pi^-$ exactly the same way as $D^0 \to K^+\pi^-$ when
one
ignores the resonant substructure. Moreover, in principle,
the detection efficiency vs
decay time may not be studied reliably by using $D^0 \to K^-\pi^+\pi^-\pi^+$
as the resonant substructure could be different for CFD and DCSD.

At the Charm 2000 workshop~\cite{Charm2000},
both E687 and E791 reported their preliminary result from part of their data.
The best upper limits on mixing should come from these two
experiments soon. Some most recent preliminary results can be found
in~\cite{Zaliznyak}, progress has been made~\cite{Peng1,Purohit1}
on the measurement of the lifetime difference
between $D^0 \to K^-\pi^+$ and $D^0 \to K^+K^-(\pi^+\pi^-)$.

\section{Comparison of Different Experiments}

\subsection{Hadronic method A}

This measurement requires:
(1) excellent vertexing capabilities, at least good enough
to see the interference structure;
(2) low background around the primary vertex.
The background level around the primary vertex
could be an important issue as the interference term in Equation 3
does peak at $t=1$.
In addition, low background around primary vertex
means that one does not suffer much from random slow pion background
and also one can measure the DCSD component at short decay times well.
This also means that good acceptance at short decay times are
very important.
These are also important for understanding DCSDs at large decay times.
The vertexing capabilities at $e^+e^-$ experiments
(${\cal L}/\sigma \sim 3$) for CLEO III and asymmetric B factories
at SLAC and KEK may be sufficient for a mixing search.
The extra path length due to the Lorentz boost,
together with the use of silicon detectors for high
resolution position measurements, have given the fixed
target experiments an advantage in vertex resolution
(typically ${\cal L}/\sigma \sim 8-10$)
over $e^+e^-$ experiments.
One major disadvantage at fixed target experiments is the poor
acceptance at short decay times.
The low background around the primary vertex at
$e^+e^-$ experiments
is a certain advantage.
It is worth pointing out here that at the
$e^+e^-$ experiments (especially at an asymmetric B factory or Z factory)
it maybe possible to use $\bar{B^0} \to D^{*+} l^-\nu $, where the
primary ($D^{*+}$ decay) vertex can be determined by the $l^-$ together with
the slow pion coming from the $D^{*+}$. In this case, the background level
around the primary vertex is intrinsically very low~\cite{Charm2000_liu}.

However, in the case of $D^0 \to K^+K^-,\pi^+\pi^-$, etc.,
the requirement on the background level around the primary
vertex is not so important. In this case,
the mixing signature is not a deviation from
a perfect exponential (again assuming CP conservation),
but rather a deviation of the
slope from $(\gamma_1 + \gamma_2)/2$.
It is worth pointing out that there are many advantages with
this method. For example, one can use Cabibbo favored decay modes,
such as $D^0 \to K^-\pi^+$, to measure the average $D^0$ decay rate
$(\gamma_1 + \gamma_2)/2$ (which is almost a pure
exponential, mixing followed by DCSD effect should be tiny, see Appendix A).
This, along with other SCSD CP even (or odd)
final states, would allow for valuable cross checks on
systematics uncertainties. In addition, since we only need to determine
the slope here, we do not need to tag the $D^0$ and do not have to
use the events close to the primary vertex.
The sensitivity of this method
depends on how well we can determine the slope difference, which needs to be
carefully studied. This is currently under study~\cite{Purohit1,Peng1}.
Roughly speaking, in the ideal case,
the sensitivity to $y$ would be $\sim 1/\sqrt{N}$, where $N$
is the number of $D^0 \to K^+K^-, \pi^+\pi^-$, etc. events,
which means that the sensitivity to mixing caused by
the decay rate difference ($\sim y^2/2$) would be close to
$\sim 1/N$.  For example, a fixed-target experiment capable of
producing $\sim 10^8$ reconstructed charm events could, in principle,
lower the sensitivity
to $\sim 10^{-5}-10^{-6}$ level for
the $y^2$ term in ${\rm R}_{\rm mixing}= (x^2 + y^2)/2$.
In reality, the sensitivity depends on many things and should be carefully
studied.

It is worth to point out that the current PDG
experimental upper limit ( $90\%$ CL)
on the life time difference is only~\cite{PDB92}
\begin{equation}
\label{detla_gamma}
\frac{|\tau_{D^0_1}-\tau_{D^0_2}|}{\tau_{D^0}}= 2~y < 17 \%
\end{equation}
This is based on the upper limit ${\rm R}_{\rm mixing} = (x^2 + y^2)/2
< 0.37 \%$, which is the E691~\cite{Browder} combined results on the
$D^0 \to K^+\pi^-$ and $D^0 \to K^+\pi^-\pi^+\pi^-$ modes
by assuming no interference
between DCSD and mixing for both modes at the same time.

\subsection{Hadronic method B}

In the near future, we should be able to have a good understanding
of DCSD~\footnote{It may be possible that
good understanding of DCSD can be reached by measuring
the pattern of $D^+$ DCSD decays where the signature is not confused
by a mixing component. It is worth pointing out that the
$D^+$ DCSD decays can be studied very well at future
fixed target experiments and B factories.}
in $D^0 \to K^+\pi^-\pi^0$, $D^0 \to K^+\pi^-\pi^+\pi^-$, etc. modes,
then method B will
become a feasible way to study mixing and the sensitivity should be
improved. Just like method A, this method requires
very good vertexing capabilities and very low
background around the primary vertex (this is even more
important than in method A, since precise knowledge of DCSD is very
important here).
In addition, this method requires that the detection efficiency (for the
mode being searched) across Dalitz plot be quite
uniform (at least the detector should have good
acceptance on the Dalitz plot at locations where DCSD and mixing
resonant substructure are different). This is necessary so that
detailed information on the resonant substructure
can be obtained in every corner on the Dalitz plot.

The excellent photon detection capabilities will allow
$e^+e^-$ experiments to study the $D^0 \to K^+\pi^-\pi^0$ mode with very low
background. From the CLEO II $D^0 \to K^+\pi^-\pi^0$
analysis~\cite{Liudpf,Thesis},
the detection efficiency across the Dalitz plot will have some
variations due to cuts needed to reduce background,
however, it is still good enough to obtain detailed information
on the resonant substructure.
Future fixed target experiments may have a good chance to
study $D^0 \to K^+\pi^-\pi^+\pi^-$ mode, since the detection efficiency
across Dalitz plot should be quite flat.
The sensitivity that each experiment can reach by using this method
depends on many things and need to be carefully studied in the future.

\subsection{Hadronic method C}

The sensitivity of this method depends crucially on the
particle identification capabilities. Since the $D^0$ is at rest,
the $K$ and $\pi$ mesons will have the same momentum, so a
doubly misidentified $D^0 \to K^-\pi^+$
($K^- \to \pi^-, \pi^+ \to K^+$) mimics a $D^0 \to K^+\pi^-$
with almost the same $D^0$ mass. It is worth pointing out here
that particle identification is not as crucial to method A
as it is to this method (C), as far as this particular background is
concerned. This is because in method A, the $D^0$
is highly boosted, and doubly misidentified $D^0 \to K^+\pi^-$ decays
will have a broad distribution in the $D^0$ mass spectrum around the
$D^0$ mass peak; this background can be kinematically rejected
with only a small reduction of the efficiency for the signal
events.

Once the sensitivity reaches ${\cal O}(10^{-5})$, one may have to
worry about other contributions, such as contributions from
continuum background, contributions from
$e^+e^- \to 2 \gamma \to D^0\bar{D^0}$ which can produce C-even
states where DCSD can contribute~\cite{Du}.

\subsection{Semi-leptonic method}

The semi-leptonic method usually suffers from large
background (except at a $\tau$ charm factory), the traditional
method of looking for like sign $\mu^+\mu^+$ or $\mu^-\mu^-$ pairs
is an example. New ideas are needed in order to improve the
sensitivity significantly. Some promising techniques
have been suggested by Morrison and
others at the Charm 2000 workshop~\cite{Charm2000}
and have been discussed in the working group~\cite{Liu1}.

The technique suggested by Morrison
is very similar to that of the hadronic method:
one uses the decay chain $D^{*+} \to D^0 \pi^+$,
instead of looking for $D^0 \to K^+\pi^-$, one can
search for $D^0 \to K^+l^-\nu$ where there
is no DCSD involved. Of course,
due to the missing neutrino, this mode usually suffers from
large background. However, for events in which the neutrino is very soft
in $D^0$ rest frame, $D^0 \to K^+l^-\nu$ is quite similar to
$D^0 \to K^+\pi^-$ kinematically. In this case, one has the same
advantages as $D^{*+} \to D^0 \pi^+$ followed by $ D^0 \to K^+\pi^-$
has. In addition, as the neutrino is soft,
the proper decay time of the $D^0$ can be reasonably estimated from
$K^+l^-$. The potential mixing signal therefore should show up
as a $t^2$ term in the proper decay time distribution.
To select the events with soft neutrino, one can require
the $K^+l^-$ mass above 1.4 GeV. This requirement will keep about
$50\%$ of the total signal.
One major background here is the random slow pion background,
as the effective mass difference width is still much larger (a factor of 10)
than $D^{*+} \to D^0 \pi^+$ followed by $ D^0 \to K^+\pi^-$.
In order to reduce this background, Morrison has suggested to look for
a lepton with the correct charge sign in the other side of the
charm decay. Another background is DCSD decay $D^0 \to K^+\pi^-$ when the
$\pi^-$ fakes a $l^-$, however, this background will only
populates at the higher end of the $K^+l^-$ mass spectrum
where the neutrino energy is almost zero. This
can be eliminated by cutting off that high end of the
$K^+l^-$ mass. In principle, this idea can be used in a
fixed target experiment as well as in a $e^+e^-$ experiment.
The sensitivity of this method depends on the
lepton fake rate (meson fakes as a lepton). One can find some detail
discussions in Morrison's Charm2000 workshop summary paper~\cite{Morrison}.

Another technique, suggested by Freyberger at CLEO, is based on
the technique which has been used by ALEPH, HRS and CLEO to extract the
number of $D^{*+} \rightarrow D^0 {\pi}_{\rm s}^+$ events.
The technique utilizes the following facts:
(1) Continuum production of $c\bar{c}$ events are jet like.
(2) The jet axis, calculated by maximizing the observed
momentum projected onto an axis, approximates the
$D^{*+}$ direction.
(3) The $D^{*+} \rightarrow D^0 {\pi}_{\rm s}^+$ decay
is a two-body process, and the small amount of energy
available means that the ${\pi}_{\rm s}^+$ is very soft,
having a transverse momentum
$p_\perp$ relative to the $D^{*+}$ direction which cannot exceed 40 MeV/c.
This low transverse momentum
provides the $D^{*+} \rightarrow D^0 {\pi}_{\rm s}^+$ signature.

The facts are used in the following way. The maximum momentum
in the lab that the ${\pi}_{\rm s}^+$ can have perpendicular to the line
of flight of the $D^{*+}$ is $40 {\rm MeV}$. One can define this quantity as
$p_{\bot}=|p_{\pi}|sin\theta_{\pi}$, where $sin\theta_{\pi}$ is the angle
between the $D^{*+}$ and the ${\pi}_{\rm s}^+$
in the lab frame, and $p_{\pi}$
is the magnitude of the ${\pi}_{\rm s}^+$ momentum. Hence, the
${\pi}_{\rm s}^+$ from $D^{*+}$ will populate the low
$p_{\bot}$ (or $sin\theta_{\pi}$) region.
The signal is enhanced if one plots
$p^2_{\bot}$ (or $sin^2\theta_{\pi}$) instead of $p_{\bot}$.
One then looks for an lepton in the jet with the correct sign, namely,
${\pi}_{\rm s}^+l^+$ right sign combination and
${\pi}_{\rm s}^+l^-$ wrong sign combination.
The signal $D^{*+} \rightarrow D^0 {\pi}_{\rm s}^+$
followed by $D^0\rightarrow K^- l^+\nu$
will peak in the low $p^2_{\bot}$ (or $sin^2\theta_{\pi}$)
region for the wrong sign events.
It is worth pointing out that one can look for a lepton
in the other side of the event to reduce background.

There are many kinds of background to this method one has to worry
about. One of the major backgrounds is fake lepton background.
For example, the decay chain
$D^{*+} \to D^0{\pi}_{\rm s}^+ \to (K^-X){\pi}_{\rm s}^+$
will also peak at the low $p^2_{\bot}$ (or $sin^2\theta_{\pi}$)
if the $K^-$ is misidentified as a $l^-$. Another major background is
probably the $\pi^0$ dalitz and $\gamma$ conversions in
$D^0 \to X \pi^0$ followed by $\pi^0 \to \gamma e^+e^-$ or
$D^0 \to X \pi^0$ followed by $\pi^0 \to \gamma \gamma$ and
then $\gamma \to e^+e^-$.
These two major backgrounds are at about $0.3\%$ level in the current
CLEO II data. Understanding these backgrounds
is the major difficulty faced by this method.
Although for CLEO III, things should improve,
it is not clear what kind of sensitivity one can
expect from this method for future experiments. Nevertheless,
it is an interesting idea and worth investigating.
In fact, this technique is currently under study at CLEO~\cite{Gibaut}.

\section{Summary}

The search for $D^0\bar{D}^0$
mixing carries a large discovery potential for new
physics since the $D^0\bar{D}^0$
mixing rate is expected to be very small in the Standard
Model.  The past decade has seen significant experimental progress in
sensitivity (from 20\% down to 0.4\%).
Despite these 18 years of effort there is still much left to be done.

As was discussed in the introduction, the observation of
$D^0 \to K^+\pi^-$ is an important step on the way to observing a potentially
small mixing signal by using this technique.
With the observation of $D^0 \to K^+\pi^-$ signal at CLEO at the level
${\rm R} = 0.0077\pm 0.0025\,({\rm stat.})\pm 0.0025\,({\rm sys.})$,
any experimenter can estimate the number of reconstructed $D^0 \to K^+\pi^-$
events their data sample will have in the future and what kind of
sensitivity their experiment could have.
Eighteen years after the search for $D^0 \to K^+\pi^-$ started,
we have finally arrived at the point where we could take advantage of
possible DCSD-mixing interference to make the mixing search easier.

In light of the CLEO's $D^0 \to K^+\pi^-$ signal,
if the mixing rate is close to that of DCSD (above $10^{-4}$)
, then it might be observed
by the year 2000 with either the hadronic or the semi-leptonic method,
either at fixed target experiments,
CLEO III, asymmetric B factories (at SLAC and KEK), or
at a $\tau$-charm factory.
If the mixing rate is indeed much smaller than DCSD,
then the hadronic method may have a better chance over the
semi-leptonic method. This is because the semi-leptonic method
usually suffers from
a large background due to the missing neutrino, while the
hadronic method does not. Moreover, the commonly believed
``annoying DCSD background'' does not necessary inherently
limit the hadronic method as
the potentially small mixing signature could show up
in the interference term.
The design of future experiments should focus on improving the
vertexing capabilities and reducing the
background level around the primary vertex, in order to
fully take advantage of having the possible DCSD and mixing interference.
In addition, we have learned that the very complication due to
the possible differences between the
resonant substructure in many DCSD and mixing decay modes
$D^0 \to K^+\pi^-(X)$ could, in principle, be turned to advantage by
providing additional information once the substructure in DCSD
is understood (the method B) and the sensitivity could be
improved significantly this way.
This means that understanding DCSD in $D$ decays could be a very important
step on the way to observe mixing. Experimenters
and theorists should work hard on this.

In the case of $D^0 \to K^+\pi^-(X)$ and $D^0 \to X^+ l^-$, we are only
measuring ${\rm R}_{\rm mixing}= (x^2 + y^2)/2 $.
Since many extensions of the Standard Model predict large
$ x={\delta m / \gamma_+}$, it is very important to measure
$x$ and $y$ separately. Fortunately, SCSD can provide us information
on $y$. This is due to the fact
that decays such as $D^0 \to K^+K^-,\pi^+\pi^-$,
occur only through definite CP eigenstate,
and this fact can be used to measure the decay rate difference
$y={\gamma_- / \gamma_+}=(\gamma_2 - \gamma_1) /(\gamma_1 + \gamma_2)$
alone. Observation of a non-zero $y$ would demonstrate mixing caused
by the decay rate difference.
This, together with the information on
${\rm R}_{\rm mixing}$ obtained from other methods, we can
in effect measure $x$. I should point out here that $x$ and $y$
are expected to be at the same level within the Standard Model, however
we do not know for sure exactly at what level since theoretical
calculations for the long distance contribution
are still plagued by large uncertainties.
Therefore, it is very important to
measure $y$ in order to understand the size of $x$
within the Standard Model, so that when $D^0\bar{D^0}$ mixing
is finally observed experimentally, we will know whether
we are seeing the Standard Model physics or new physics beyond
the Standard Model.

In this sense, it is best
to think of the quest to observe mixing (new physics) as a program rather than
a single effort.

\section*{Acknowledgement}

Much of my knowledge on this subject has been gained by
having worked closely with Hitoshi Yamamoto over the past four years,
I would like to express my sincere gratitude to him.
Besides, I would like to thank the organizers of the workshop,
especially Jose Repond, for his patient and encouragement.
This work is supported by Robert H. Dicke Fellowship Foundation at
Princeton University.

\appendix

\pagebreak

\section{Appendix A
--- The Time Dependence of $D^0 \to K^+\pi^-$}

Note that Appendices A, B and C are from reference~\cite{Thesis}.

\subsection{The time-dependent effect}

A pure $D^0$ state generated at $t=0$ could decay to $K^+\pi^-$ state either
by $D^0-\bar{D^0}$ mixing or by DCSD, and the two amplitudes may interfere.
Following the notation in~\cite{Thesis},
the time evolutions of $|D_1>$ and $|D_2>$ are given by
\begin{equation}
\label{d1d2}
|D_i(t)> = e_i |D_i>, \;\;\; e_i = e^{-im_it-\frac{\gamma_i}{2}t}, \;\;(i=1,2)
\end{equation}
with $m_i,\gamma_i~(i=1,2)$ being the masses and decay rates of the two CP
eigenstates,
The mass eigenstates $|D_1>$ and $|D_2>$ are given by
\begin{equation}
\label{mass_eigen3}
|D_1> = p~|D^0> + q~|\bar{D^0}>
\end{equation}
\begin{equation}
\label{mass_eigen3}
|D_2> = p~|D^0> - q~|\bar{D^0}>
\end{equation}
where
\begin{equation}
\label{p_q_epsilon}
\frac{p}{q}=\frac{1+\epsilon}{1-\epsilon}
\end{equation}
Under the phase convention $|\bar{D^0}> = CP |D^0>$,
a state that is purely $|D^0>$ ($|\bar{D^0>}$) prepared by the strong
interaction at $t=0$
will evolve to $|D^0_{\rm phys}(t)>$ ($|\bar{D^0}_{\rm phys}(t)>$)
\begin{equation}
\label{d0t1}
|D^0_{\rm phys}(t)> = \frac{1}{2}\left[~(e_1+e_2) |D^0>
+ \frac{q}{p} (e_1 - e_2) |\bar{D^0}>~\right]
\end{equation}
\begin{equation}
\label{d0bart1}
|\bar{D^0}_{\rm phys}(t)> = \frac{1}{2}\left[~\frac{p}{q}(e_1 - e_2) |D^0>
+ (e_1 + e_2) |\bar{D^0}>~\right]
\end{equation}
Let us define
$a(f) = Amp(D^0 \to f)$, $\bar{a}(f) = Amp(\bar{D^0} \to f)$
with $ \rho (f)= a(f)/\bar{a}(f)$;
and $a(\bar{f}) = Amp(D^0 \to \bar{f})$,
$\bar{a}(\bar{f}) = Amp(\bar{D^0} \to \bar{f})$
with $\bar{\rho}(\bar{f})= \bar{a}(\bar{f})/a(\bar{f})$.
Now the decay amplitude for states initially pure $D^0$ ($\bar{D^0}$) to decay
to $f$ ($\bar{f}$)
is given by (with $f=K^+\pi^-$ and define $|\bar{f}> \equiv CP |f>$)
\begin{equation}
\label{mixingamp0_kpi}
Amp (|D^0_{\rm phys}(t)> \to f) =
\frac{1}{2}~\bar{a}(f)\left[~\rho (f) (e_1+e_2)
+ \frac{q}{p}~(e_1-e_2)~\right]
\end{equation}
\begin{equation}
\label{mixingamp0bar_kpi}
Amp (|\bar{D^0}_{\rm phys}(t)> \to \bar{f}) =
\frac{1}{2}~a(\bar{f})\left[~\frac{p}{q}~(e_1-e_2)
+ \bar{\rho}(\bar{f})~(e_1+e_2)~\right]
\end{equation}
Therefore, we have
\begin{equation}
\label{mixingamp0}
I(~|D^0_{\rm phys}(t)> \to f~) =
\frac{|\bar{a}(f)|^2}{4}~|\frac{q}{p}|^2~|~(e_1-e_2) +
\frac{p}{q}\rho (f)~(e_1+e_2)~|^2
\end{equation}
\begin{equation}
\label{mixingamp0bar}
I(~|\bar{D^0}_{\rm phys}(t)> \to \bar{f}~)
= \frac{|a(\bar{f})|^2}{4}~|\frac{p}{q}|^2~|~(e_1-e_2) +
\frac{q}{p}\bar{\rho} (\bar{f})~(e_1+e_2)~|^2
\end{equation}
Note that neither $\frac{p}{q}$ nor $\rho (f)$ is an observable by itself.
The flavor eigenstates $D^0$ and $\bar{D^0}$ are defined by the strong
interaction
only, thus the relative phase between $D^0$ and $\bar{D^0}$ are undetermined:
$|\bar{D^0}> \equiv e^{-i\alpha} CP |D^0>$ (we have set $\alpha =0$ for
convenience).
The phase transformation $|\bar{D^0}> \to e^{i\alpha}|\bar{D^0}>$ leads to
$\frac{p}{q} \to e^{i\alpha}\frac{p}{q}$ and $\rho(f) \to e^{-i\alpha}\rho(f)$.
However, under the phase transformation,
$\frac{p}{q}\rho(f) \to \frac{p}{q}\rho(f)$ remains unchanged which means that
$\frac{p}{q}\rho(f)$ is an observable and we define $\eta \equiv
\frac{p}{q}\rho(f)$
($\bar{\eta} \equiv \frac{q}{p}\bar{\rho}(\bar{f})$ ).

We also have
\begin{equation}
\label{e1pluse2}
e_1+e_2 = 2~e^{-im_+t-\frac{\gamma_+}{2}t}~\cosh(\frac{i}{2}\delta
m~t+\frac{1}{4}\delta{\gamma}~t)
\end{equation}
and
\begin{equation}
\label{e1minse2}
e_1-e_2 = 2~e^{-im_+t-\frac{\gamma_+}{2}t}~\sinh(\frac{i}{2}\delta
mt+\frac{1}{4}\delta{\gamma}t)
\end{equation}
where $m_+ = (m_1+m_2)/2$,
$\delta m = m_2-m_1$, $\delta{\gamma} = \gamma_2-\gamma_1$ and
$\gamma_+ = (\gamma_1+\gamma_2)/2$ which is the average
$D^0$ decay rate.
With $\xi = \frac{1}{2}(~i\delta m~+~\frac{1}{2}\delta{\gamma}~)t$, we get
\begin{equation}
\label{mixingrate}
I(~|D^0_{\rm phys}(t)> \to f~) = |\bar{a}(f)|^2~|\frac{q}{p}|^2~|
{}~\sinh(\xi) + \eta~\cosh(\xi)~
|^2~e^{-\gamma_+ t}
\end{equation}
\begin{equation}
\label{mixingratebar}
I(~\bar{D^0}_{\rm phys}(t)> \to \bar{f}~) = |a(\bar{f})|^2~|\frac{p}{q}|^2~|
{}~\sinh(\xi) + \bar{\eta}~\cosh(\xi)~
|^2~e^{-\gamma_+ t}.
\end{equation}

Note for $\delta mt,\delta{\gamma}t << 1$ and for small $|\eta|$ (
in the case of $f=K^+\pi^-$), equation~\ref{mixingrate} becomes
\begin{equation}
\label{mixingrate_kpi}
I(~|D^0_{\rm phys}(t)> \to f~) =
|\bar{a}(f)|^2~|\frac{q}{p}|^2~|~\frac{1}{2}(i\delta m +
\frac{1}{2}\delta \gamma )~t + \eta~|^2~ e^{-\gamma_+t}
\end{equation}
or
\begin{eqnarray}
\label{mixingrate2}
\lefteqn{I(|D^0_{\rm phys}(t)> \to f) =
\frac{|\bar{a}(f)|^2}{4}~|\frac{q}{p}|^2~\times} \nonumber \\
 & & {\left[ 4\eta^2
+ (~(\delta m)^2 + \frac{(\delta{\gamma})^2}{4}~)t^2
+ 2 (~\delta{\gamma}{\rm Re}\eta + 2\delta m~{\rm
Im}\eta)~t~\right]e^{-\gamma_+t} }
\end{eqnarray}
and equation~\ref{mixingratebar} becomes
\begin{eqnarray}
\label{mixingrate2bar}
\lefteqn{I(|\bar{D^0}_{\rm phys}(t)> \to \bar{f}) =
\frac{|a(\bar{f})|^2}{4}~|\frac{p}{q}|^2~\times} \nonumber \\
 & & {\left[ 4\bar{\eta}^2
+ (~(\delta m)^2 + \frac{(\delta{\gamma})^2}{4}~)t^2
+ 2 (~\delta{\gamma}{\rm Re}\bar{\eta} + 2\delta m~{\rm
Im}\bar{\eta})~t~\right]e^{-\gamma_+t} }
\end{eqnarray}
Note the difference between equation~\ref{mixingrate2} and~\ref{mixingrate2bar}
is the indication of CP violation. These equations are in the form which
people usually use. Next I will write the equations in a different form,
which I believe is more convenient for discussion here.

Now let us look at the same equation in a different way. For convenience,
let us assume CP conservation so that we have $|\frac{p}{q}| = 1$
and $|\bar{a}(f)|=|a(\bar{f})|=|a|$.
Equation~\ref{mixingrate_kpi} can be written in the form
\begin{equation}
\label{mixingamp0}
I(|D^0_{\rm phys}(t)> \to f) = |a|^2~|~\frac{ix+y}{2}~t +
\eta~|^2~e^{-t}
\end{equation}
where now the time t is measured in unit of average $D^0$ lifetime
($1/\gamma_+$),
and
$$ x={\delta m \over \gamma_+},\qquad
   y={\gamma_- \over \gamma_+} $$
$$ \delta m = m_2 - m_1,\qquad
   \gamma_\pm = {\gamma_2\pm \gamma_1 \over 2} $$
Note again that we have assumed a small mixing; namely,
$\delta m, \gamma_- \ll \gamma_+$ or $x,y\ll 1$,
which means we have ${\rm R}_{\rm mixing} = (x^2 + y^2)/2$.
In addition, let's define
${\rm R}_{\rm DCSD}=|\eta|^2$.
Note when $|\frac{p}{q}| \sim 1$, ${\rm R}_{\rm DCSD}=|\eta|^2
=|\rho|^2$, which is the natural
definition of ${\rm R}_{\rm DCSD}$ (see equation~\ref{rho_dcsd}).
Now we have
\begin{equation}
\label{mixingamp}
I(|D^0_{\rm phys}(t)> \to f) = |a|^2~|~\sqrt{{\rm R}_{\rm mixing}/2}\:\; t +
\sqrt{{\rm R}_{\rm DCSD}}\:\; e^{i\phi}~|^2~e^{-t}
\end{equation}
where
\begin{equation}
\label{mixingphase}
\phi=Arg(ix+y)-Arg(\eta)
\end{equation}
which is an unknown phase, and this gives:
\begin{eqnarray}
\label{time_dep_kpi}
\lefteqn{I(|D^0_{\rm phys}(t)> \to f) = |a|^2~\times} \nonumber \\
  & & {\left[{\rm R}_{\rm DCSD} +
\sqrt{2{\rm R}_{\rm mixing}{\rm R}_{\rm DCSD}}~t~\cos\phi +
\frac{1}{2}{\rm R}_{\rm mixing}\:\; t^2~\right]~e^{-t}}
\end{eqnarray}

Note the unknown phase $\phi=Arg(ix+y)-Arg(\eta)$ depends not only on the
mixing parameters, which are the sign and relative size of
$x$ ($\delta m$) and $y$ ($\delta \gamma$),
but also on the relative phase between DCSD $D^0 \to K^+\pi^-$
and Cabibbo favored decay $\bar{D^0} \to K^+\pi^-$ which is due to
final state interaction.
We can define
$\phi_{\rm mixing}=Arg(ix+y)$ with
$\phi_{\rm DCSD}=Arg(\eta)=Arg(\frac{p}{q}) + Arg(\rho(f))$.
Note $x=\sqrt{2{\rm R}_{\rm mixing}}~\sin\phi_{\rm mixing}$ and
$y=\sqrt{2{\rm R}_{\rm mixing}}~\cos\phi_{\rm mixing}$.
By using equation~\ref{time_dep_kpi}, one can only measure
${\rm R}_{\rm DCSD}$, ${\rm R}_{\rm mixing}$ and $\cos\phi$ (up to a sign
for $\phi$),
but not $\phi_{\rm DCSD}$ which depends on final state interaction
and is unknown, neither $\phi_{\rm mixing}$ which depends on the sign and
relative size of $x$ ($\delta m$) and $y$ ($\delta \gamma$).
Note that since $\eta \to \eta$ under the phase transformation
$|\bar{D^0}> \to e^{i\alpha}|\bar{D^0}>$, $\phi_{\rm DCSD}=Arg(\eta)$
is a physical parameter.
In order to measure $\phi_{\rm DCSD}$ and
$\phi_{\rm mixing}$, one can use $D^0 \to K^+K^-$ etc. to measure
$y$ ($\delta \gamma$) as discussed in method A. Together with
${\rm R}_{\rm mixing}$ and $\cos\phi$ (up to a sign for $\phi$)
measured using equation~\ref{time_dep_kpi},
one can thus in effect measure $x$ or $\delta m$ (up to a sign)
, therefore
$\phi_{\rm mixing}$ (up to a sign) and
$\phi_{\rm DCSD} (=\pm |\phi_{\rm mixing}| \pm |\phi|$).

It has recently been argued by Browder and
Pakvasa~\cite{Tom} that it is possible that
$\phi_{\rm DCSD}$ can be calculated. Although at present the calculations
are purely phenomenological and are plagued by large uncertainties,
it is worth investigating since it
would be very helpful if $\phi_{\rm DCSD}$ can be calculate
theoretically in a reliable way.

As discussed above, the difference between
equation~\ref{mixingrate2} and~\ref{mixingrate2bar}
is the indication of CP violation.
Equation~\ref{mixingrate2} and~\ref{mixingrate2bar}
can be written in the form:
\begin{eqnarray}
\label{time_dep_kpi_cp}
\lefteqn{I(|D^0_{\rm phys}(t)> \to f) =
|\bar{a}(f)|^2~|\frac{q}{p}|^2~\times} \nonumber \\
 & & {\left[{\rm R}_{\rm DCSD} +
\sqrt{2{\rm R}_{\rm mixing}{\rm R}_{\rm DCSD}}~t~\cos\phi +
\frac{1}{2}{\rm R}_{\rm mixing}\:\; t^2~\right]~e^{-t} }
\end{eqnarray}
and
\begin{eqnarray}
\label{time_dep_kpi_cp_bar}
\lefteqn{I(|\bar{D^0}_{\rm phys}(t)> \to \bar{f}) =
|a(\bar{f})|^2~|\frac{p}{q}|^2~\times} \nonumber \\
 & & {\left[\overline{{\rm R}}_{\rm DCSD} +
\sqrt{2{\rm R}_{\rm mixing}\overline{{\rm R}}_{\rm DCSD}}~t~\cos\bar{\phi} +
\frac{1}{2}{\rm R}_{\rm mixing}\:\; t^2~\right]~e^{-t} }
\end{eqnarray}
where $\overline{{\rm R}}_{\rm DCSD}=|\bar{\eta}|^2$ and
$\bar{\phi}=Arg(ix+y)-Arg(\bar{\eta})$. Note $\bar{\phi}$ is
different from $\phi$ because $Arg(\bar{\eta}) = Arg(\frac{q}{p})
+ Arg(\bar{\rho}(\bar{f}))$ where $Arg(\frac{q}{p})=-Arg(\frac{p}{q})$.
For convenience, let us ignore the CP violation in the
decay amplitude. Thus the difference between the interference
terms (that is, the interference phase $\phi$ and $\bar{\phi}$)
in equation~\ref{time_dep_kpi_cp} and~\ref{time_dep_kpi_cp_bar}
would be the indication of CP violation.

It has been recently argued~\cite{CPinmixing1,CPinmixing2}
that CP violation effect
should not be ignored here, since new physics which can
introduce large mixing rate may often involve significant
CP violation.
In the SU(3) limit, the quark model and factorization gives
$Arg(\rho(f))=0$. In this special case, and also assuming
that $x>>y$ (large mixing caused by new physics) which means
$\phi_{\rm mixing}=\pi/2$, the interference term
now simply becomes $\pm ~xsin(2\phi_{M})~t$ where
$\frac{q}{p}=e^{-2i\phi_{M}}$. Under those specific assumptions,
the interferece term is odd with respect to CP, thus CP violating.
This is the case discussed recently in~\cite{CPinmixing2}.
In general, due to final state interaction,
we have~\cite{Yamamoto} $Arg(\rho(f)) \neq 0$.
In addition, in the Standard Model, $x$ and $y$ are
expected to be at the same level. But we do not know
exactly at what level since theoretical predictions
are still plagued by large uncertainties. For instance,
if ${\rm R}_{\rm mixing} \sim
10^{-4}$ in the Standard Model~\cite{Bigi5}, then
$y$ should be somewhere between $10^{-3}$ to $10^{-2}$.
Not to mention the current upper limit on $y$
is at $10^{-1}$ level (this is why measurement
on $y$ by using $D^0 \to K^+K^-$ etc is so important,
as discussed before).
Therefore, unless new physics introduces large $x$ far above
$10^{-3}$, one should use
equation~\ref{time_dep_kpi_cp} and~\ref{time_dep_kpi_cp_bar}
or equation~\ref{mixingrate2} and ~\ref{mixingrate2bar}
to fit data, instead of assuming $x>>y$ (or $\phi_{\rm mixing}=\pi/2$).
It is worth to point out here that we are looking for a tiny
effect, the possible contribution due to the decay rate difference $y$
should not be ignored unless $x>>y$ is experimentally confirmed.

As discussed in method A,
a small mixing signature could be enhanced by DCSD
through interference at lower
decay times, compared to the case without interference.
This really depends on the actual value of
$\cos\phi$. If nature is unkind, the sign and relative size of
$x$ ($\delta m$) and $y$ ($\delta \gamma$), and
$\phi_{\rm DCSD}$ caused by final state interaction are
such that $\cos\phi =\cos(\phi_{\rm mixing} - \phi_{\rm DCSD})
=0$, then we will not have this advantage by using this
technique. In this case, we may have to use
$D^0 \to K^+\pi^-\pi^0$ (method B)
where $\cos\phi$ strongly depends on the
location on the Dalitz plot, and cannot be zero everywhere
on the Dalitz plot. In general, $D^0 \to K^+\pi^-\pi^0$ may
provide much more information on mixing and CP violation
than $D^0 \to K^+\pi^-$
does. We will discuss this point in more detail
in appendix B.

It is worth to take a look at the time dependence of the right sign
decay $\bar{D^0} \to K^+\pi^-$ here. The decay amplitude for states
initially pure $\bar{D^0}$ to decay to $f$ is given by:
\begin{equation}
\label{mixingamp0bar_kpi_r}
Amp (|\bar{D^0}_{\rm phys}(t)> \to f) =
\frac{1}{2}~\bar{a}(f)\left[~\frac{p}{q}~\rho(f)(e_1-e_2)
+ (e_1+e_2)~\right]
\end{equation}
Therefore we have
\begin{equation}
\label{mixingamp0bar_r}
I(~|\bar{D^0}_{\rm phys}(t)> \to f~)
= \frac{|\bar{a}(f)|^2}{4}~|~\eta~(e_1-e_2) + (e_1+e_2)~|^2
\end{equation}
which is
\begin{equation}
\label{mixingratebar_r1}
I(~\bar{D^0}_{\rm phys}(t)> \to f~) = |\bar{a}(f)|^2~|
\eta~\sinh(\xi) + \cosh(\xi)~
|^2~e^{-\gamma_+ t}.
\end{equation}
Similarly, we have
\begin{equation}
\label{mixingrate_r1}
I(~D^0_{\rm phys}(t)> \to \bar{f}~) = |a(\bar{f})|^2~|
\bar{\eta}~\sinh(\xi) + \cosh(\xi)~
|^2~e^{-\gamma_+ t}.
\end{equation}
Note for $\delta mt,\delta{\gamma}t << 1$,
equations~\ref{mixingratebar_r1} and ~\ref{mixingrate_r1} become
\begin{equation}
\label{mixingratebar_r2}
I(~|\bar{D^0}_{\rm phys}(t)> \to f~) =
|\bar{a}(f)|^2~|~\frac{1}{2}~\eta~(i\delta m +
\frac{1}{2}\delta \gamma )~t + 1~|^2~ e^{-\gamma_+t}
\end{equation}
and
\begin{equation}
\label{mixingrate_r2}
I(~|D^0_{\rm phys}(t)> \to \bar{f}~) =
|a(\bar{f})|^2~|~\frac{1}{2}~\bar{\eta}~(i\delta m +
\frac{1}{2}\delta \gamma )~t + 1~|^2~ e^{-\gamma_+t}
\end{equation}

Assuming CP conservation, Equation~\ref{mixingratebar_r2}
can also be written in the form:
\begin{equation}
\label{mixingamp_r}
I(|\bar{D^0}_{\rm phys}(t)> \to f) =
|a|^2~|~\sqrt{{\rm R}_{\rm DCSD}{\rm R}_{\rm mixing}/2}\:\; t e^{i\phi^{'}}
+1~|^2~e^{-t}
\end{equation}
where
\begin{equation}
\label{mixingphase_r}
\phi^{'}=Arg(ix+y)+Arg(\eta)
\end{equation}
which is an unknown phase, and this gives:
\begin{eqnarray}
\label{time_dep_kpi_r}
\lefteqn{I(|\bar{D^0}_{\rm phys}(t)> \to f) = |a|^2~\times} \nonumber \\
  & & {\left[~1+
\sqrt{2{\rm R}_{\rm mixing}{\rm R}_{\rm DCSD}}~t~\cos\phi^{'} +
\frac{1}{2}{\rm R}_{\rm mixing}{\rm R}_{\rm DCSD} \:\; t^2~\right]~e^{-t}}
\end{eqnarray}

{}From equation~\ref{time_dep_kpi_r}
one can see that the mixing term is further suppressed by
${\rm R}_{\rm DCSD}$ due to the fact that $\bar{D^0} \to K^+\pi^-$
can occur through mixing followed by DCSD $D^0 \to K^+\pi^-$.
The interference term is similar to that of equation~\ref{time_dep_kpi},
but the interference phase $\phi^{'}$ is different from the $\phi$
in equation~\ref{time_dep_kpi}. Since the first term is the dominated one,
equation~\ref{time_dep_kpi_r} is essentially the same as
$I(|\bar{D^0}_{\rm phys}(t)> \to f) = |a|^2~e^{-t}$.

\subsection{The time-integrated effect}

It is interesting to take a look at the time integrated effect. Let us define
\begin{equation}
\label{time_int_kpi_define}
{\rm R}=\frac{\int_{0}^{\infty} I(|D^0_{\rm phys}(t)> \to f)\,dt}
{\int_{0}^{\infty} I(|\bar{D^0}_{\rm phys}(t)> \to f)\,dt}
\end{equation}
and we have
\begin{equation}
\label{time_int_kpi}
{\rm R}= {\rm R}_{\rm DCSD}
+\sqrt{2{\rm R}_{\rm mixing}{\rm R}_{\rm DCSD}}\; \cos\phi
+ {\rm R}_{\rm mixing}.
\end{equation}

In the special case when $|\cos\phi|=1$, equation~\ref{time_int_kpi}
can be written in the form
\begin{equation}
\label{time_int_kpi_sp}
(~{\rm R}_{\rm DCSD}-{\rm R}~)^2
+ (~{\rm R}_{\rm mixing}-{\rm R}~)^2 = {\rm R}^2.
\end{equation}

If one can measure ${\rm R}$ and ${\rm R}_{\rm DCSD}$ precisely,
then ${\rm R}_{\rm mixing}$ can only have two possible values
${\rm R}_{\rm mixing} = {\rm R} \pm \sqrt{2{\rm R}_{\rm DCSD}({\rm R}-{\rm
R}_{\rm DCSD}/2)}$
where the sign $\pm$ corresponds to $\cos\phi= \mp 1$.
For example, if one finds ${\rm R} = {\rm R}_{\rm DCSD} = {\rm R_0}$,
then ${\rm R}_{\rm mixing}$ can be either 0 or $2{\rm R_0}$.

In general, we do not know $\cos\phi$, so we cannot determine
${\rm R}_{\rm mixing}$ this way. However, there are still useful
information in the time integrated effect.  For instance,
one can compare ${\rm R}(t \leq 0.2)$, which is
measured by using events that have decay times $t \leq 0.2$,
to the ${\rm R}$ for all decay times.
At short decay times where mixing has not yet fully developed,
the wrong sign events will be almost pure DCSD (see Section 6.2.1),
which means we have ${\rm R}_{\rm DCSD}={\rm R}(t \leq 0.2)$.
Equation~\ref{time_int_kpi} tells us that
${\rm R}$ will be different from ${\rm R}_{\rm DCSD}$
by
$\sqrt{2{\rm R}_{\rm mixing}{\rm R}_{\rm DCSD}}\; \cos\phi
+{\rm R}_{\rm mixing}$.
Compared to ${\rm R}_{\rm DCSD}$, the fraction of
the difference is
$\sqrt{2{\rm R}_{\rm mixing}/{\rm R}_{\rm DCSD}}\; \cos\phi
+{\rm R}_{\rm mixing}/{\rm R}_{\rm DCSD}$.
For ${\rm R}_{\rm mixing} \sim 10^{-4}$,
${\rm R}_{\rm DCSD} \sim 7.7 \times 10^{-3}$ and $|\cos\phi|=1$,
this corresponds to a $\pm \sim 16\%$ change, which could be
measurable with reasonable amount of data.
Note here that one can only measure
$\Delta {\rm R} = {\rm R} - {\rm R}_{\rm DCSD}$.
Observation of a non-zero $\Delta {\rm R}$ would demonstrate
the existence of mixing.
In addition, a precise measurement on
${\rm R}_{\rm DCSD}$ using wrong sign events at short decay times
would help the time-dependent analysis using equation~\ref{time_dep_kpi}.
For instance, it could be helpful to fixed target experiment
if ${\rm R}_{\rm DCSD}$ has been already measured very well at other
experiments.
This is because that at fixed target experiment, the detection efficiency
at short
decay times is very low, leading to a poor constraint on the large DCSD
term and
making the time-dependent analysis very difficult.

As the detection
efficiency cancels out in the wrong sign vs. right sign
ratio~\footnote{This is because we will combine $D^{*+}$ and
$D^{*-}$ decays, and we have assumed CP conservation.
}, the method discussed
above would be useful when one does not know
the detection efficiency vs. decay time therefore cannot
perform a reliable time-dependent analysis.
However, in the case of $D^0 \to K^+\pi^-$, one can
always use the right sign events $D^0 \to K^-\pi^+$ to
study the detection efficiency. Nevertheless, this should
still be a useful cross-check for the time-dependent analysis.
For example, one can use the fitted  $\cos\phi$ value
from the time-dependent analysis to determine
${\rm R}_{\rm mixing}$
in the time-integrated analysis. This would be a valuable
cross-check for the measured value of
${\rm R}_{\rm mixing}$ for the time-dependent analysis.

If one wants to measure CP violation, one could then
compare equation~\ref{time_int_kpi} for $D^0 \to K^+\pi^-$
and $\bar{D^0} \to K^-\pi^+$ separately. Note that in this
case one does not need to measure ${\rm R}_{\rm DCSD}$ which is
good for experiments without decay time information at all.
However, since one has to measure
$D^0 \to K^+\pi^-$
and $\bar{D^0} \to K^-\pi^+$ separately the detection efficiency
would not drop out completely. This means it could be quite
difficult since we are looking for a tiny effect here.


\section{Appendix B --- The Time Dependence of
$D^0 \to K^+\pi^-\pi^0$}

\subsection{The time-dependent Dalitz analysis}

In the case of $D^0 \to K^+\pi^-\pi^0$, the situation is very similar to
that of $D^0 \to K^+\pi^-$. For a given location $p$
on the $D^0 \to K^+\pi^-\pi^0$
Dalitz plot,
the decay amplitude for states initially pure $D^0$ to decay to $f$
is given by (with $f=K^+\pi^-\pi^0$)
\begin{equation}
\label{mixingamp0_kpipi0}
Amp (|D^0_{\rm phys}(t)> \to f)(p) = \frac{1}{2}~\bar{a}(f,p)\left[~\rho (f,p)
(e_1+e_2)
+ \frac{q}{p}~(e_1-e_2)~\right]
\end{equation}
where $a(f,p) = Amp(D^0 \to f)(p)$, $\bar{a}(f,p) = Amp(\bar{D^0} \to f)(p)$
and with $ \rho (f,p)= a(f,p)/\bar{a}(f,p)$.
Note that mathematically, equation~\ref{mixingamp0_kpipi0} is
exactly the same as equation~\ref{mixingamp0_kpi}.
Therefore, for a given location $p$ on the Dalitz plot, we have
\begin{equation}
\label{time_dep_kpipi0}
I(~|D^0_{\rm phys}(t)> \to f~)(p)= |\bar{a}(f,p)|^2~|\frac{q}{p}|^2~|
{}~\sinh(\xi) + \eta(p)~\cosh(\xi)~ |^2~e^{-\gamma_+ t}
\end{equation}
where again $\xi = \frac{1}{2}(~i\delta m~+~\frac{1}{2}\delta{\gamma}~)~t$,
and $\eta(p) = \frac{p}{q}\rho(f,p)$ which now depends on the location
$p$ on the Dalitz plot. Note in general, $|\eta(p)| \ll 1$ does
not hold. For instance, $|\eta(p)|$ could be large at
locations on the Dalitz plot where DCSD amplitude is large due to
constructive interference among DCSD submodes,
while CFD amplitude is small due to destructive
interference among CFD submodes. Nevertheless, as long as
$|\eta(p)| \ll min((\frac{\delta m}{\gamma_+})^{-1},
(\frac{\delta \gamma}{2\gamma_+})^{-1})$,
for $\delta mt,\delta{\gamma}t << 1$ we still have
\begin{equation}
\label{mixingrate_kpipi0}
I(~|D^0_{\rm phys}(t)> \to f~) (p) =
|\bar{a}(f,p)|^2~|\frac{q}{p}|^2~|~\frac{1}{2}(i\delta m +
\frac{1}{2}\delta \gamma )~t + \eta(p)~|^2~ e^{-\gamma_+t}
\end{equation}
just like in the case of $D^0 \to K^+\pi^-$, this can be written
in the form
\begin{equation}
\label{time_dep_kpipi01}
I(~|D^0_{\rm phys}(t)> \to f~) (p)= |a(p)|^2~|~\sqrt{{\rm R}_{\rm
mixing}/2}\:\; t +
\sqrt{{\rm R}_{\rm DCSD}(p)}\:\; e^{i\phi(p)}~|^2~e^{-t}
\end{equation}
where we have assumed CP conservation, with $|\bar{a}(f,p)|^2=|a(\bar{f},p)|^2
\equiv |a(p)|^2$, and
${\rm R}_{\rm DCSD}(p) \equiv |\eta(p)|^2$ as the definition of
${\rm R}_{\rm DCSD}(p)$ at location $p$ on the Dalitz plot. Now
we have
\begin{eqnarray}
\label{time_dep_kpipi0}
\lefteqn{I(~|D^0_{\rm phys}(t)> \to f~)~(p) = |a(p)|^2~\times} \nonumber \\
 & & {\left[{\rm R}_{\rm DCSD}(p) +
\sqrt{2{\rm R}_{\rm mixing}{\rm R}_{\rm DCSD}(p)}~t~\cos\phi(p) +
\frac{1}{2}{\rm R}_{\rm mixing}\:\; t^2~\right]~e^{-t} }
\end{eqnarray}
Note again
the unknown phase $\phi(p)=Arg(ix+y)-Arg(\eta(p))$ depends not only on the
mixing parameters, which are the sign and relative size of
$x$ ($\delta m$) and $y$ ($\delta \gamma$),
but also on the relative phase between DCSD $D^0 \to K^+\pi^-\pi^0$
and Cabibbo favored decay $D^0 \to K^-\pi^+\pi^0$ at location $p$ on the
Dalitz plot.
As in the case of $D^0 \to K^+\pi^-$, we can define
$$\phi_{\rm mixing}=Arg(ix+y)$$
$$\phi_{\rm DCSD}(p)=Arg(\eta(p)) = Arg(\frac{p}{q}) + Arg(~\rho(f,p)~)$$
$$\phi(p)=\phi_{\rm mixing}-\phi_{\rm DCSD}(p)$$.

Let $|b(p)|^2 = |a(f,p)|^2 = n_{\rm D}(p)$ which is the
true yield density on the Dalitz plot for direct
DCSD decay $D^0 \to K^+\pi^-\pi^0$, and $|a(p)|^2 = n_{\rm C}(p)$
which is the true yield density for the direct CFD decay
$D^0 \to K^-\pi^+\pi^0$,
equation~\ref{time_dep_kpipi01} can be written in the form
\begin{equation}
\label{time_dep_kpipi02}
I(~|D^0_{\rm phys}(t)> \to f~)~(p) = |~\sqrt{n_{\rm C}(p)}~\sqrt{{\rm R}_{\rm
mixing}/2}\:\; t +
\sqrt{n_{\rm D}(p)}\:\; e^{i\phi(p)}~|^2~e^{-t},
\end{equation}
which gives
\begin{eqnarray}
\label{time_dep_kpipi02}
\lefteqn{I(~|D^0_{\rm phys}(t)> \to f~)~(p) = }  \nonumber \\
   & & {\left[~n_{\rm D}(p)
+ \sqrt{2{\rm R}_{\rm mixing}~n_{\rm D}(p)~n_{\rm C}(p)}~\cos\phi(p)~t
+ \frac{1}{2}~n_{\rm C}(p)~{\rm R}_{\rm mixing}~t^2\right]~e^{-t}. }
\end{eqnarray}
where $n_{\rm C}(p)$ and
$n_{\rm D}(p)$ can be written in the
form
\begin{equation}
\label{density_cfd}
n_{\rm C}(p) \:\propto \:  |~f_{1}\:e^{i\phi_{1}}A_{3b} +
f_{2}\:e^{i\phi_{2}}BW_{\rho^{+}}(p) + f_{3}\:e^{i\phi_{3}}BW_{K^{*-}}(p)
+f_{4}\:e^{i\phi_{4}}BW_{\bar{K}^{*0}}(p)~|^2,
\end{equation}
and
\begin{equation}
\label{density_dcsd}
n_{\rm D}(p) \:\propto \:  |~g_{1}\:e^{i\theta_{1}}A_{3b} +
g_{2}\:e^{i\theta_{2}}BW_{\rho^{-}} + g_{3}\:e^{i\theta_{3}}BW_{K^{*+}}
+g_{4}\:e^{i\theta_{4}}BW_{K^{*0}}~|^2,
\end{equation}
where $f_{i}$ ($g_{i}$) are the relative amplitudes for each component for
CFD (DCSD)
and $\phi_{i}$
($\theta_{i}$) are the relative
interference phases between each submode for CFD (DCSD).
$A_{3b}$ is the S-wave three-body
decay amplitude, which is assumed to be flat across the Dalitz plot. The
various
$BW(p)$ terms are Breit-Wigner amplitudes for the $K^{*}\pi$ and
$K\rho$ sub-reactions,
which describe the strong resonances and decay angular momentum conservation:
$BW_{R} \:\propto \: \frac{\cos\theta_{R}}{M_{ij}-M_{R}-i\Gamma_{R}/2}$ where
$M_{R}$ and $\Gamma_{R}$ are the mass and width of the $M_{ij}$
resonance ($K^{*}$ or $\rho$), and $\theta_{R}$ is the helicity angle
of the resonance.

The idea of the time-dependent Dalitz analysis is that,
in principle, one could perform a multi-dimensional
fit to the data by using the information on the mass difference,
$D^0$ mass, proper decay time $t$ and the yield density on the
wrong sign Dalitz plot. Of course, in order to do this, one needs
a very large amount of clean data - a formidable experimental challenge!
Lest any reader's despair, we would like to remind the reader that
CP violation was first discovered 31 years ago in $K_{\rm L} \to \pi^+\pi^-$
with only 45 events with $S/N \sim 1$, and now we are talking about $10 \times
10^6$
$K_{\rm L} \to \pi^+\pi^-$ events at KTeV with a expected background
level at $0.03\%$~\cite{Briere}!

As we have pointed out, the extra information on the
resonant substructure will, in principle, put a much better
constraint on the amount of mixing (compare to that of
$D^0 \to K^+\pi^-$). A Monte Carlo study is currently underway and
will be described elsewhere.
Here we will just make some qualitative remarks:

(a) The amplitudes $f_i$ and relative interference phases $\phi_i$
in equation~\ref{density_cfd}, thus $n_{\rm C}(p)$,
can be measured very well for
the Cabibbo favored decay $D^0 \to K^-\pi^+\pi^0$. Note that
only relative phases are measurable, so out of the four phases
there is one unknown (denoted $\phi_0$).

(b) As pointed out in method A,
in principle, one can use wrong sign samples at short
decay time (which is almost pure DCSD) to study the resonant
substructure of the DCSD decays. This means
the amplitudes $g_i$ and relative interference phases $\theta_i$
in equation~\ref{density_dcsd}, thus $n_{\rm D}(p)$,
can also be measured for
the DCSD $D^0 \to K^+\pi^-\pi^0$. Note again that
only relative phases are measurable, so out of the four phases
there is one unknown (denoted $\theta_0$).

(c) Now let us look at $\eta(p) = \frac{p}{q}\rho(f,p)$ where
$ \rho (f,p)= a(f,p)/\bar{a}(f,p)$.
Assuming CP conservation, we have $\eta(p) = \frac{p}{q}b(p)/a(p)$
where $|\frac{p}{q}|=1$ and the phase of $\frac{p}{q}$
is unphysical as it depends on
phase convention (we can set it to zero for convenience).
Once we know $f_i (g_i)$ and $\phi_i(\theta_i)$ as can be measured in
(a) and (b), then $\phi_{\rm DCSD}(p) = Arg(\eta(p))$ will only depend
on $\phi_0 - \theta_0$ for a given location $p$ on Dalitz plot.
The sign and size of $\phi_0 - \theta_0$ is due to final
state interaction and is unknown.
At different locations, the phase $\phi_{\rm DCSD}(p)$ will be
different as it strongly depends on the location due to
the phases of various Breit-Wigner amplitudes
($BW_{R} \:\propto \: \frac{cos\theta_{R}}{M_{ij}-M_{R}-i\Gamma_{R}/2}$).
Nevertheless, in principle, for given CFD and
DCSD resonant substructure, there is only one unknown phase
which is $\phi_0 - \theta_0$,
and $\phi_0 - \theta_0$ does not depend on location $p$
on the Dalitz plot.

(d) The phase $\phi_{\rm mixing} = Arg(ix + y)$ depends
on the sign and relative size of $x$ and $y$. For given
$x$ and $y$, $\phi_{\rm mixing}$ is fixed.

(e) Therefore, for given CFD and DCSD resonant substructure and
mixing parameters ($x$ and $y$), the mixing-DCSD interference
phase $\phi(p)=Arg(ix+y)-Arg(\eta(p))$ will only depend on
$\phi_0 - \theta_0$ at location $p$. For a given
$\phi_0 - \theta_0$, one would expect that $\cos\phi(p)$
could have any value between $[-1, 1]$, depending on the
location $p$. Therefore, one can perform
a Monte Carlo study by changing $\phi_0 - \theta_0$ and
look at the wrong sign Dalitz plot at certain decay time $t$.
This Monte Carlo study is currently underway.

(f) It is interesting to look at the locations on the Dalitz plot
where $\cos\phi(p)=-1$ (maximal destructive mixing-DCSD interference).
As we have discussed in method A, at decay time
$t_0 = \sqrt{2{\rm R}_{\rm DCSD}(p)/{\rm R}_{\rm mixing}}$,
we have $|~\sqrt{{\rm R}_{\rm mixing}/2}\:\; t +
\sqrt{{\rm R}_{\rm DCSD}(p)}\:\; e^{i\phi(p)}~| =0$.
This means the mixing-DCSD interference would dig a ``hole''
on the Dalitz plot at time $t_0$ at that location. Since
$t_0 \propto \sqrt{{\rm R}_{\rm DCSD}(p)}=\sqrt{n_{\rm D}(p)/n_{\rm C}(p)}$,
the ``holes'' would show up earlier (in decay time) at locations where
${\rm R}_{\rm DCSD}(p)$ is smaller. Imagine that someone watches the
Dalitz plot as the decay time ``goes by'', this person
would expect to see ``holes'' moving
from locations with $\cos\phi(p)=-1$ and smaller
${\rm R}_{\rm DCSD}(p)$ toward locations with
$\cos\phi(p)=-1$ and larger ${\rm R}_{\rm DCSD}(p)$.
The existence of the ``moving holes'' on the Dalitz plot would
be clear evidence for mixing.

(g) In practice, one can only expect to find the ``holes''
at earlier decay times, since at longer decay times there are
not many events left. As already pointed out in (f), the
``holes'' showing up at early decay times have small
${\rm R}_{\rm DCSD}(p)$.
In general, small ${\rm R}_{\rm DCSD}(p)$ occurs at locations where
the DCSD amplitude is small due to destructive interference among
DCSD submodes while
the CFD amplitude is large due to constructive interference among
CFD submodes.
This means that it could be difficult to see the ``holes'' since
there is not many wrong sign events to begin with at these locations,
unless one has a very large amount of clean data.

(h) There are also extra information at locations where $\cos\phi(p) \neq -1$.
Therefore, the best way would be to perform a multi-dimensional fit to data by
using
the information on the mass difference, $D^0$ mass, proper decay time $t$ and
the
yield density on the wrong sign Dalitz plot.

In the discussions above, we have assumed CP conservation.
As we discussed in appendix A, CP violation effect could be important here.
It is interesting to point out here that the extra information
on the resonant substructure could also provide additional information
on CP violation when one compares the difference between the
resonant substructure of $D^0 \to K^+\pi^-\pi^0$ and
$\bar{D^0} \to K^-\pi^+\pi^0$.
As discussed in (c), assuming CP conservation,
once we know $f_i (g_i)$ and $\phi_i(\theta_i)$ as can be measured in
(a) and (b), then $\phi_{\rm DCSD}(p) = Arg(\eta(p))$ will only depend
on $\phi_0 - \theta_0$ for a given location $p$ on Dalitz plot.
With CP violation, then $Arg(\eta(p))$ will also depends on
$Arg(\frac{p}{q})$. This is for $D^0 \to K^+\pi^-\pi^0$. For
$\bar{D^0} \to K^-\pi^+\pi^0$, $Arg(\bar{\eta}(p))$ will
depends on $Arg(\frac{q}{p}) = -Arg(\frac{p}{q})$. This means
the interference phase of
$\bar{D^0} \to K^-\pi^+\pi^0$ will be shifted by
$2~Arg(\frac{p}{q}$ ) everywhere on the Dalitz plot,
compared to that of $D^0 \to K^+\pi^-\pi^0$. This would
lead to a totally different resonant substructure.
For example, similar to what has been discussed in (f),
if the person sees a ``hole'' at location $p$ and decay time $t$
on the $D^0 \to K^+\pi^-\pi^0$ Dalitz plot, but could not find a
``hole'' on the $\bar{D^0} \to K^-\pi^+\pi^0$ Dalitz plot
at the same location $p$ and decay time $t$, that would be indication
of CP violation. This will be studied by Monte Carlo soon.
In any case, with enough data in the future, one should fit
$D^0 \to K^+\pi^-\pi^0$ and $\bar{D^0} \to K^-\pi^+\pi^0$
separately, or study the difference between the two.

As in appendix A, it is also worth to take a look at
the time dependence of the right sign decay
$\bar{D^0} \to K^+\pi^-\pi^0$ here.
Assuming CP conservation, just as Equation~\ref{time_dep_kpi_r}
we have for $\bar{D^0} \to K^+\pi^-\pi^0$
\begin{eqnarray}
\label{time_dep_kpipi0_r}
\lefteqn{I(|\bar{D^0}_{\rm phys}(t)> \to f) = |a(p)|^2~\times}  \nonumber \\
  & & {\left[~1+
\sqrt{2{\rm R}_{\rm mixing}{\rm R}_{\rm DCSD}(p)}~t~\cos\phi^{'}(p) +
\frac{1}{2}{\rm R}_{\rm mixing}{\rm R}_{\rm DCSD}(p) \:\; t^2~\right]~e^{-t} }
\end{eqnarray}
where
\begin{equation}
\label{mixingphase_r}
\phi^{'}(p)=Arg(ix+y)+Arg(\eta(p))
\end{equation}

Unlike in the case of $\bar{D^0} \to K^+\pi^-$ where
${\rm R}_{\rm DCSD} \sim 0.77\%$ which is very
small, in the case of $\bar{D^0} \to K^+\pi^-\pi^0$
${\rm R}_{\rm DCSD}(p)$ depends on location $p$ and,
in principle, could be quite large at certain locations.
If mixing is indeed large, then the interference
term could be important at locations where
${\rm R}_{\rm DCSD}(p)$ is also large. For instance,
if ${\rm R}_{\rm mixing} \sim 10^{-3}$ and
for locations where ${\rm R}_{\rm DCSD}(p) \sim 10$,
then the interference term
$\sqrt{2{\rm R}_{\rm mixing}{\rm R}_{\rm DCSD}(p)}$ would be
at $10\%$ level. If mixing is indeed small, then the
second and third terms can be ignored, and equation~\ref{time_dep_kpipi0_r}
simply becomes
$I(|\bar{D^0}_{\rm phys}(t)> \to f) = |a(p)|^2~e^{-t}$.

\subsection{The time-integrated Dalitz analysis}
One alternative would be to study the time integrated Dalitz plot. Now
equation~\ref{time_dep_kpipi0} becomes:
\begin{eqnarray}
\label{time_int_kpipi0}
\lefteqn{I(|D^0_{\rm phys}> \to f)~(p)~\propto |a(p)|^2~\times} \nonumber \\
  & & {\left[~{\rm R}_{\rm DCSD}(p)
+\sqrt{2{\rm R}_{\rm mixing}{\rm R}_{\rm DCSD}(p)}\; \cos\phi(p)
+ {\rm R}_{\rm mixing}~\right]}.
\end{eqnarray}
As we have already discussed, at short decay times, mixing has not yet fully
developed,
the resonant substructure will be almost due to DCSD alone.
Therefore, in principle, one can measure ${\rm R}_{\rm DCSD}(p)$
or $n_{\rm D}(p)$ this way. Note again that
$\phi(p)=Arg(ix+y)-Arg(\eta(p))$, which can be written in the form
\begin{equation}
\label{phase}
\phi(p)=\Phi(p) + \phi
\end{equation}
where $\Phi(p)$ is the part which depends on the location $p$ on the Dalitz
plot,
and $\phi$ is the part which does not.

Once we know the resonant substructure for
CFD and DCSD, we know ${\rm R}_{\rm DCSD}(p)$ and $\Phi(p)$.
There are only two parameters in equation~\ref{time_int_kpipi0}
which are unknown: (1) ${\rm R}_{\rm mixing}$ which is what we are
trying to measure; (2) $\phi$ which depends on
$Arg (ix +y)$ and $\phi_0 - \theta_0$ which is the overall relative phase
between DCSD and CFD.
Therefore, in principle, one could also perform a multi-dimensional fit to data
by using
the information on the mass difference, $D^0$ mass, and the
yield density on the wrong sign Dalitz plot to measure these two unknown
parameters. However, in practice, since we expect that the interference term
is much larger than the mixing term, it may not be easy to separate
${\rm R}_{\rm mixing}$ and the phase $\phi$ effects since they enter the
interference
term as $\sqrt{{\rm R}_{\rm mixing}}\; \cos(\Phi(p)+ \phi)$.

One could also compare
the Dalitz plot for decay times $t \leq 0.2$
(which is almost due to pure DCSD)
to the Dalitz plot for all decay times (or for $t > 0.2$).
At longer decay times, for given
${\rm R}_{\rm mixing}$, the resonant substructure would
be modified according to $\cos\phi(p)$. Since $\cos\phi(p)$
only depends on the location $p$, but not on the decay time $t$,
the interference effect will not be integrated out.
Equation~\ref{time_int_kpipi0} tells us that at location $p$
the wrong sign resonant substructure will be changed
by approximately
$\sqrt{2{\rm R}_{\rm mixing}{\rm R}_{\rm DCSD}(p)}\; \cos\phi(p)$,
compared to ${\rm R}_{\rm DCSD}(p)$ measured at short decay times.
The fraction of change is
$\sqrt{2{\rm R}_{\rm mixing}/{\rm R}_{\rm DCSD}(p)}\; \cos\phi(p)$.
For ${\rm R}_{\rm mixing} \sim 10^{-4}$,
${\rm R}_{\rm DCSD}(p) \sim 10^{-2}$ and $|\cos\phi(p)|=1$,
this corresponds to $\pm \sim 10\%$ change, which could be
measurable with reasonable amount of data. Note that since
${\rm R}_{\rm DCSD}(p)$ is different at different location on
Dalitz plot, the fraction of change will also be different
at different location for given mixing rate.
A clear difference between
the Dalitz plot for decay times $t \leq 0.2$
(which is almost due to pure DCSD)
to the Dalitz plot for all decay times (or for $t > 0.2$)
would be clear evidence for the signature of
mixing, although the information on the size of
${\rm R}_{\rm mixing}$ could be somewhat washed out by the unknown phases.

Note that the time-integrated Dalitz plot for
$D^0 \to K^+\pi^-\pi^0$ and $\bar{D^0} \to K^-\pi^+\pi^0$
can also be used to study CP violation as discussed in
Section 7.1. This is simply because the interference
effect on the Dalitz plot will not be integrated out.

\subsection{The time-dependent Non-Dalitz analysis}
Another possible alternative would be to study the time-dependent
effect after integrating over the whole Dalitz plot, or put it another
way, one could ignore the resonant substructure and treat it simply exactly as
$D^0 \to K^+\pi^-$ time-dependent analysis. In fact, this is what E691 did to
$D^0 \to K^+\pi^-\pi^+\pi^-$ (E791 and E687 are doing the same thing now).
With limited data size, this is a natural thing to try. However one must
be very careful here since one cannot treat multi-body decays exactly
as $D^0 \to K^+\pi^-$. We will discuss this point below.

After integrating equation~\ref{time_dep_kpipi02} over the whole
Dalitz plot, the decay rate $I(~|D^0_{\rm phys}(t)> \to f~)$
is now proportional to
\begin{eqnarray}
\label{time_dep_kpipi02_int}
\lefteqn{\int n_{\rm C}(p)~da~\times
\left[~\frac{\int n_{\rm D}(p) da}{\int n_{\rm C}(p) da}  \right.} \nonumber \\
 & & { \left. + \sqrt{2{\rm R}_{\rm mixing}}~t~\frac{\int \sqrt{~n_{\rm
D}(p)~n_{\rm C}(p)}~\cos\phi(p)~da}
{\int n_{\rm C}(p) da}
+ \frac{1}{2}~{\rm R}_{\rm mixing}~t^2\right]~e^{-t}. }
\end{eqnarray}
Now one can define
\begin{equation}
\label{total_r}
{\cal R}_{\rm DCSD} = \frac{\int n_{\rm D}(p) da}{\int n_{\rm C}(p) da}
\end{equation}
which is what one could measure if there is no mixing (or use events at short
decay times). However, in general, the interference term is not proportional
to $\sqrt{{\cal R}_{\rm DCSD}}$. Therefore, one cannot treat $D^0 \to
K^+\pi^-\pi^0$
(thus also $D^0 \to K^+\pi^-\pi^+\pi^-$) exactly the same way as $D^0 \to
K^+\pi^-$.
Note that since the interference term in $D^0 \to K^+\pi^-$ and
$D^0 \to K^+\pi^-\pi^0$ are totally different, one cannot combine the
two results to set upper limit, as E691~\cite{Browder} did
to $D^0 \to K^+\pi^-$ and $D^0 \to K^+\pi^-\pi^+\pi^-$  by assuming the
interference phase to be zero in both cases.

In principle, if the following inequality holds~\footnote{
This inequality should hold in general. Intuitively,
inequality~\ref{compare} is similar to inequality
$\sum_{i=1}^{n}x_{i}~\sum_{j=1}^{n}y_{j} \geq (~\sum_{i=1}^{n}\sqrt{x_{i}y_{i}}
{}~\cos\phi_{i}~)^2$, which is trivial to prove.
}
\begin{equation}
\label{compare}
\sqrt{\int n_{\rm D}(p)da~ \int n_{\rm C}(p) da} \geq
\int \sqrt{~n_{\rm D}(p)~n_{\rm C}(p)}~|\cos\phi(p)|~da
\end{equation}
one can force
\begin{equation}
\label{inter_def}
\frac{\int \sqrt{~n_{\rm D}(p)~n_{\rm C}(p)}~\cos\phi(p)~da}
{\int n_{\rm C}(p) da} \equiv \sqrt{{\cal R}_{\rm DCSD}}~\cos\Psi
\end{equation}
With this definition, equation~\ref{time_dep_kpipi02} can be written in the
form
\begin{equation}
\label{dalitz_int}
\int n_{\rm C}(p)~da \left[~ {\cal R}_{\rm DCSD}
+ \sqrt{2{\rm R}_{\rm mixing}~{\cal R}_{\rm DCSD}}~t~\cos\Psi
+ \frac{1}{2}~{\rm R}_{\rm mixing}~t^2\right]~e^{-t}.
\end{equation}

Mathematically, equation~\ref{dalitz_int} is similar to that of
$D^0 \to K^+\pi^-$, i.e. equation~\ref{time_int_kpi}.
But the ``interference'' phase $\Psi$ has a totally
different meaning since $\cos\Psi$ strongly depends on the actual value of
$\int \sqrt{~n_{\rm D}(p)~n_{\rm C}(p)}~\cos\phi(p)~da$. For example,
in the extreme case where for any given value of
$\sqrt{~n_{\rm D}(p)~n_{\rm C}(p)}$, if $\cos\phi(p)$ has an equal chance
of being negative and positive (with the same size), then the interference
effect could be completely integrated out, leading to $\cos\Psi =0$.
In general, since $\cos\phi(p)$ can be positive and negative, the
interference effect could be greatly reduced. Therefore, this technique does
not necessary have the advantage as in the $D^0 \to K^+\pi^-$ time-dependent
analysis.

\subsection{The time-integrated Non-Dalitz analysis}
Another possible alternative would be to study the effect after integrating
over decay time and also integrating over the whole Dalitz plot, this is in
fact
what has been done at CLEO~\cite{Liudpf,Thesis}.

Integrating equation~\ref{time_dep_kpipi02_int} over all decay times,
and normalizing to the corresponding right sign samples, we get
\begin{equation}
\label{int_all}
{\cal R} =~\frac{\int n_{\rm D}(p)da}{\int n_{\rm C}(p) da}
+ \sqrt{2{\rm R}_{\rm mixing}}~\frac{\int \sqrt{~n_{\rm D}(p)~n_{\rm
C}(p)}~\cos\phi(p)~da}
{\int n_{\rm C}(p) da}
+ {\rm R}_{\rm mixing}~.
\end{equation}
Note the information on mixing is essentially lost due to the lack of
precise knowledge on $n_{\rm D}(p)$ and $\cos\phi(p)$.

Again if the inequality~\ref{compare} holds, equation~\ref{int_all}
can be written in the form
\begin{equation}
\label{int_all1}
{\cal R} = {\cal R}_{\rm DCSD}
+ \sqrt{2{\rm R}_{\rm mixing}~{\cal R}_{\rm DCSD}}~\cos\Psi
+ {\rm R}_{\rm mixing}~.
\end{equation}

\subsection{Summary of Appendix B}
Although it is true that more-than-two body $D^0$ hadronic decays, such as
$D^0 \to K^+\pi^-\pi^0$, are very complicated due to the possible difference
in the resonant substructure of the DCSD and CFD (mixing) decays, this
unique attribute could, in principle, provide additional information
which could allow one to distinguish DCSD and mixing, and to study CP
violation in the future.
The best way would be to perform a time-dependent Dalitz analysis, that is,
one can perform a multi-dimensional fit to the data by using
the information on $\Delta M$, $M (D^0)$, proper decay time $t$, and the
yield density on Dalitz plot $n_w(p,t)$. The extra information
on the resonant substructure will, in principle, put a much
better constraint on the amount of mixing and CP violation.
For experiments without decay time information, one could perform
a time-integrated Dalitz analysis. This could be a feasible way
to observe mixing, but may not be a good way to measure or constrain
mixing rate. With limited data size, one could in principle,
perform a time-dependent non-Dalitz analysis, as one would do to
$D^0 \to K^+\pi^-$. However, care must be taken
here since one cannot treat multi-body decays the same as
$D^0 \to K^+\pi^-$. In addition, since the extra information on
the resonant substructure has been integrated out, the
time-dependent non-Dalitz analysis would not have
much advantage. Without decay time information and with limited data
size, one could only perform a time-integrated non-Dalitz analysis as
has been done in~\cite{Liudpf,Thesis}.

As one can see, precise knowledge of the resonant substructure for DCSD
is important to this technique.
This means that understanding
DCSD in $D$ decays could be a very important step on the way to observing
mixing and CP violation using this technique.
In principle, one can use the wrong sign sample at very low decay
times (which is almost pure DCSD) to study the resonant
substructure of the DCSD decays.
It is also worth pointing out that it
may be possible that a good understanding of DCSD could be reached
by measuring the pattern
of $D^+$ DCSD decays where the signature is not confused by a potential
mixing component. In the near future, we should have a good understanding
of DCSD decays and this method could become a feasible way to search
for $D^0\bar{D}^0$ mixing and CP violation.

\section{Appendix C --The Interference Between Mixing and DCSD in $B^0$ Case}

As we have discussed, the interference between $D^0\bar{D^0}$
mixing and DCSD can be used to study
$D^0\bar{D^0}$ mixing and CP violation. It is interesting
to point out that the interference between mixing and DCSD
also occurs in $B^0\bar{B^0}$ system. We will use
$B^0_{d} \to D^+\pi^-$ as an example (for convenience, let's also call
it DCSD, as it is doubly Cabibbo suppressed relative to $B^0_{d} \to
D^-\pi^+$).
In this case, mixing is quite large and can be well measured
while DCSD is small and unknown.
The small signature of DCSD would
mostly show up in the interference term. But here we are not interested
in measuring mixing nor measuring DCSD, what is interesting here is to
measure CP violation.

Equation~\ref{mixingrate} and~\ref{mixingratebar}
also apply to $B^0 \to D^+\pi^-$. We have (with $f=D^+\pi^-$)
\begin{equation}
\label{btodpirate}
I(~|B^0_{\rm phys}(t)> \to f~) = |\bar{a}(f)|^2~|\frac{q}{p}|^2~|
{}~\sinh(\xi) + \eta~\cosh(\xi)~
|^2~e^{-\gamma_+ t}
\end{equation}
where $\xi = \frac{1}{2}~(i\delta m+\frac{1}{2}~\delta{\gamma}~)~t$.
We have defined $\eta= \frac{p}{q}\rho(f)$, with
$ \rho (f)= a(f)/\bar{a}(f)$ and
$a(f) = Amp(B^0 \to f)$, $\bar{a}(f) = Amp(\bar{B^0} \to f)$.

We also have
\begin{equation}
\label{btodpiratebar}
I(~\bar{B^0}_{\rm phys}(t)> \to \bar{f}~) = |a(\bar{f})|^2~|\frac{p}{q}|^2~|
{}~\sinh(\xi) + \bar{\eta}~\cosh(\xi)~
|^2~e^{-\gamma_+ t}
\end{equation}
where we have defined
$\bar{\eta}= \frac{q}{p}\bar{\rho}(\bar{f})$, with
$\rho(\bar{f})= \bar{a}(\bar{f})/a(\bar{f})$,
$a(\bar{f}) = Amp(B^0 \to \bar{f})$ and
$\bar{a}(\bar{f}) = Amp(\bar{B^0} \to \bar{f})$.

In the case of $B^0_{d}\bar{B^0}_{d}$ mixing, $\delta \gamma \ll \delta m$,
which means we have $\xi = \frac{i}{2}\delta m$. Note that
$\sinh(ix)=i~sin(x)$ and $\cosh(ix)= cos(x)$, so that
equation~\ref{btodpirate} and~\ref{btodpiratebar}
becomes
\begin{eqnarray}
\label{btodpirate1}
\lefteqn{I(~|B^0_{\rm phys}(t)> \to f~) =
|\bar{a}(f)|^2~|\frac{q}{p}|^2~\times} \nonumber \\
 & & {\left[
|\eta|^2 + (1-\eta|^2)~\sin^2(\frac{1}{2}\delta m~t) +
Im (\eta) \sin(\delta m~t)~\right]~e^{-\gamma_+ t}, }
\end{eqnarray}
and
\begin{eqnarray}
\label{btodpiratebar1}
\lefteqn{I(~\bar{B^0}_{\rm phys}(t)> \to \bar{f}~) =
|a(\bar{f})|^2~|\frac{p}{q}|^2~\times }  \nonumber \\
  & & {\left[
|\bar{\eta}|^2 + (1-|\bar{\eta}|^2)\sin^2(\frac{1}{2}\delta m~t) +
Im (\bar{\eta}) \sin(\delta m~t)~\right]~e^{-\gamma_+ t}. }
\end{eqnarray}
For simplicity, let us first neglect the phase difference
between $B^0 \to D^+\pi^-$ and $\bar{B^0} \to D^+\pi^-$
caused by final state interaction.
Define $\eta = |\eta|~e^{i\Phi}$
with ${\rm R}_{\rm DCSD}= |\eta|^2$ and
$\bar{{\rm R}}_{\rm DCSD}= |\bar{\eta}|^2$,
assuming $|\frac{p}{q}| \sim 1$,
we have $\bar{\eta}=\eta^{*}=|\eta|~e^{-i\Phi}$.

\begin{eqnarray}
\label{btodpirate2}
\lefteqn{I(~|B^0_{\rm phys}(t)> \to f~) = |\bar{a}(f)|^2~\times}  \nonumber \\
  & & {\left[
{\rm R}_{\rm DCSD} + (1-{\rm R}_{\rm DCSD})\sin^2(\frac{1}{2}x~t) +
\sqrt{{\rm R}_{\rm DCSD}}~\sin{\Phi}~\sin(xt)~\right]~e^{-t} , }
\end{eqnarray}
and
\begin{eqnarray}
\label{btodpiratebar2}
\lefteqn{I(~\bar{B^0}_{\rm phys}(t)> \to \bar{f}~) =
|a(\bar{f})|^2~\times}   \nonumber \\
  & & {\left[
{\rm R}_{\rm DCSD} + (1-{\rm R}_{\rm DCSD})\sin^2(\frac{1}{2}x~t) -
\sqrt{{\rm R}_{\rm DCSD}}~\sin{\Phi}~\sin(xt)~\right]~e^{-t} ,}
\end{eqnarray}
where now the time t is measured in unit of average $B^0$
lifetime ($1/\gamma_+$), and $ x={\delta m \over \gamma_+}$.

\begin{figure}[p]
\unitlength 1in
\begin{picture}(6.5,6)(0,0)
\put(-.45,-1.25){\psfig{width=6.96in,height=9.5in,%
file=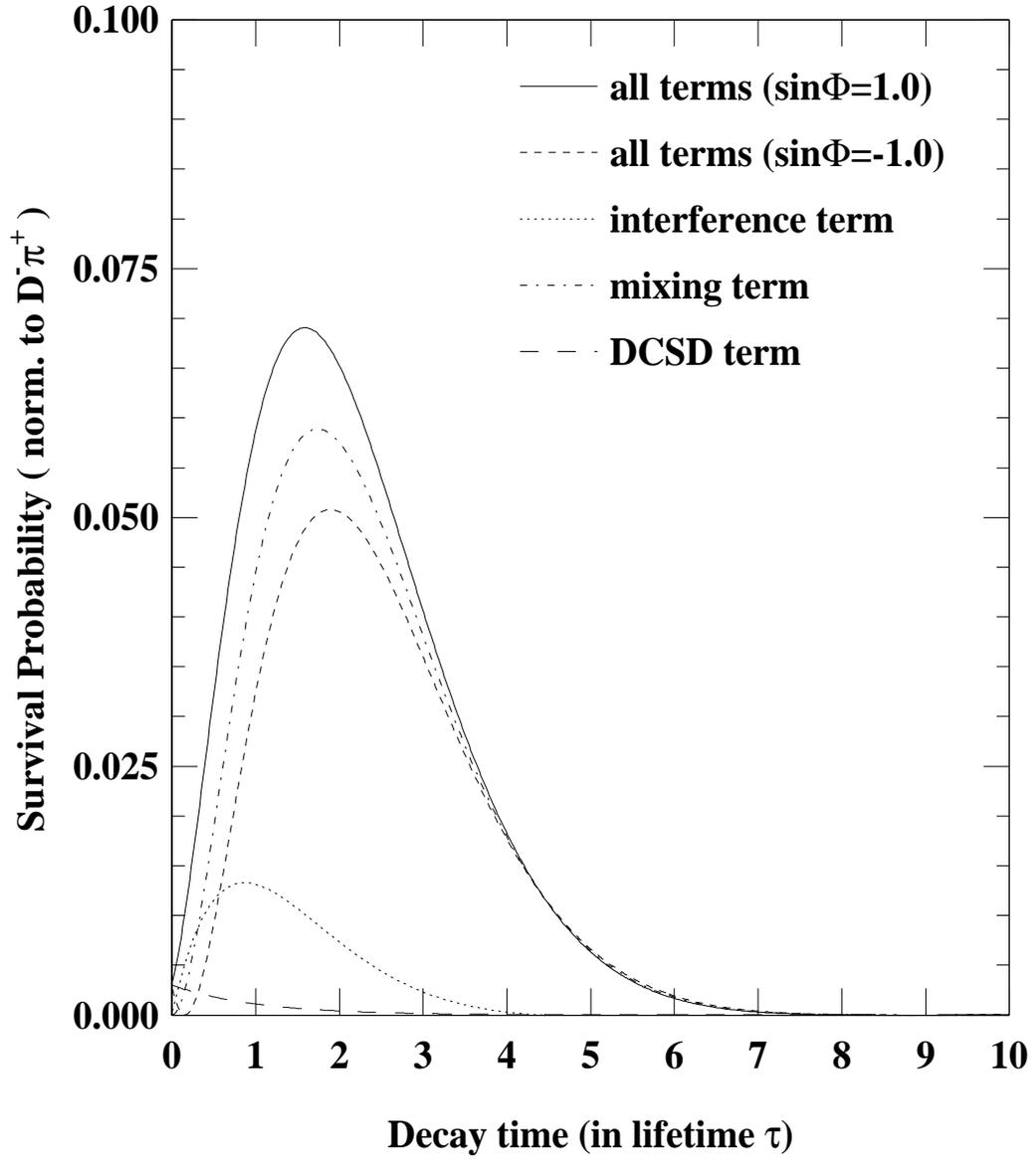}}
\end{picture}
\caption
{The decay time dependence of $I(~|B^0_{\rm phys}(t)> \to D^+\pi^-~)$
($\sin\Phi=1.0$)
and
$I(~\bar{B^0}_{\rm phys}(t)> \to  D^-\pi^+~)$ ($\sin\Phi=-1.0$)}
\label{btodpi_cp_sin1}
\end{figure}

\begin{figure}[p]
\unitlength 1in
\begin{picture}(6.5,6)(0,0)
\put(-.45,-1.25){\psfig{width=6.96in,height=9.5in,%
file=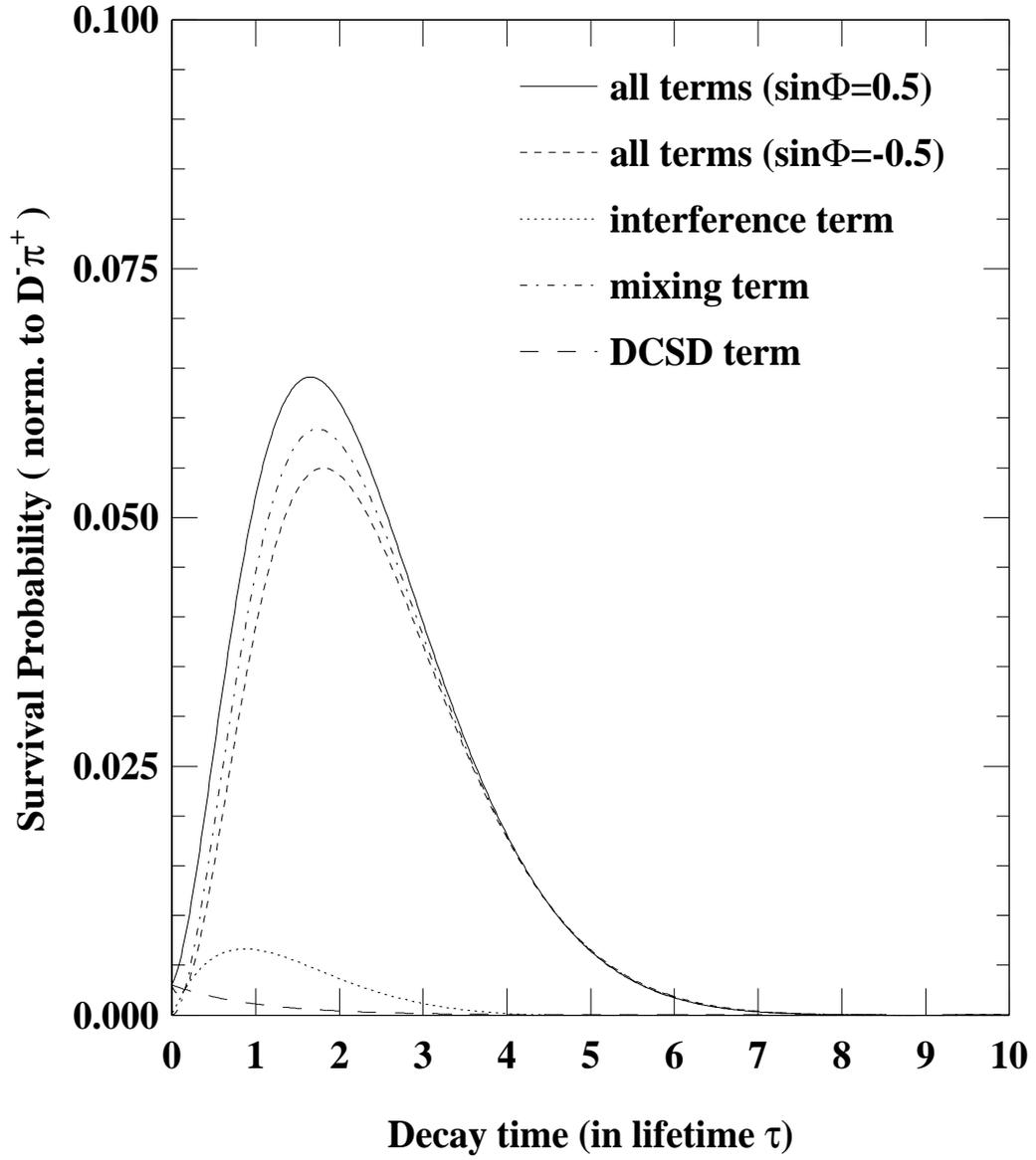}}
\end{picture}
\caption
{The decay time dependence of $I(~|B^0_{\rm phys}(t)> \to D^+\pi^-~)$
($\sin\Phi=0.5$)
and
$I(~\bar{B^0}_{\rm phys}(t)> \to  D^-\pi^+~)$ ($\sin\Phi=-0.5$)}
\label{btodpi_cp_sin05}
\end{figure}

We do not know ${\rm R}_{\rm DCSD}$ for $B^0 \to D^+\pi^-$, but
one would expect very roughly
${\rm R}_{\rm DCSD} \sim tan^4{\theta_{C}} \sim 0.3\%$.
Note again we have neglected the phase difference between
$B^0 \to D^+\pi^-$ and $\bar{B^0} \to D^+\pi^-$ caused by final
state interaction. Assuming $\sin{\Phi}=1.0$ (corresponds to
a maximal CP violation), with $x=0.71$ and ${\rm R}_{\rm DCSD}=0.3\%$,
and also set $|a(\bar{f})|^2 = |\bar{a}(f)|^2=1$,
Figure~\ref{btodpi_cp_sin1} shows the decay time dependence of
equation~\ref{btodpirate2} and ~\ref{btodpiratebar2}, together
with each term alone.
One can see that there is a sizeable interference between
$B^0\bar{B^0}$ mixing and small DCSD.
More importantly,
there is a clear difference between
$I(~|B^0_{\rm phys}(t)> \to D^+\pi^-~)$ and
$I(~\bar{B^0}_{\rm phys}(t)> \to  D^-\pi^+~)$.
Note that the mixing term peaks at $t \simeq 2$, where
mixing is fully developed and the maximal
asymmetry occurs.
Figure~\ref{btodpi_cp_sin05} shows the same thing, but with
$\sin{\Phi}=0.5$.

It is interesting to take a look at the time-integrated decay rate:
\begin{eqnarray}
\label{btodpirate3}
\lefteqn{\Gamma(~|B^0_{\rm phys}> \to f~) =
|\bar{a}(f)|^2~\times}  \nonumber \\
   & & {\left[
{\rm R}_{\rm DCSD} + (1-{\rm R}_{\rm DCSD})\frac{x^2}{2(1+x^2)}
+ \frac{x}{1+x^2}\sqrt{{\rm R}_{\rm DCSD}}~\sin{\Phi}~\right],}
\end{eqnarray}
and
\begin{eqnarray}
\label{btodpiratebar3}
\lefteqn{\Gamma(~|\bar{B^0}_{\rm phys}> \to \bar{f}~) =
|a(\bar{f})|^2~\times}  \nonumber \\
 & & {\left[
{\rm R}_{\rm DCSD} + (1-{\rm R}_{\rm DCSD})\frac{x^2}{2(1+x^2)}
- \frac{x}{1+x^2}\sqrt{{\rm R}_{\rm DCSD}}~\sin{\Phi}~\right].}
\end{eqnarray}
Therefore the time-integrated asymmetry is (with
${\rm R}_{\rm DCSD} \ll \frac{x^2}{2(1+x^2)}$)
\begin{eqnarray}
\label{btodpiratebar}
\lefteqn{<Asym> = } \nonumber \\
 & & {\frac{\Gamma(~|B^0_{\rm phys}> \to D^+\pi^-~)
-~\Gamma(~|\bar{B^0}_{\rm phys}> \to D^-\pi^+~)}
{\Gamma(~|B^0_{\rm phys}> \to D^+\pi^-~)
+~\Gamma(~|\bar{B^0}_{\rm phys}> \to D^-\pi^+~)} }  \nonumber \\
  & & {\simeq \frac{2}{x}\sqrt{{\rm R}_{\rm DCSD}}~\sin{\Phi}. }
\end{eqnarray}
With $x=0.71$ and ${\rm R}_{\rm DCSD} \sim 0.3\%$,
we have $<Asym> = 15\% \times~\sin{\Phi}$ which is
quite large. Similar conclusion can be made on
the lepton-tagged $B^0 \to D^+\pi^-$ on $\Upsilon(4S)$.

Of course, we have neglected the phase difference between
$B^0 \to D^+\pi^-$ and $\bar{B^0} \to D^+\pi^-$ caused
by final state interaction. In general, however, one cannot
ignore the phase difference and things would be more complicated.

After I thought about this, I found that this mode has been
discussed in the past. For people who are interested in
this mode, relevant references can be found in ~\cite{Sachs} to~\cite{xing}.

It is also worth to point out that there could be a possibility of using
multi-body decays such as $B^0 \to D^+\pi^-\pi^0$
to study CP violation, similar to
the case of $D^0 \to K^+\pi^-\pi^0$ where one
could use the extra information on the resonant
substructure.

\pagebreak

     \def\pl#1#2#3#4{#1, Phys.\ Lett.\ {\bf#2}, #3 (#4)}
     \def\plb#1#2#3#4{#1, Phys.\ Lett.\ {\bf#2B}, #3 (#4)}
     \def\pr#1#2#3#4{#1, Phys.\ Rev.\ {\bf#2}, #3 (#4)}
     \def\prd#1#2#3#4{#1, Phys.\ Rev.\ D {\bf#2}, #3 (#4)}
     \def\prl#1#2#3#4{#1, Phys.\ Rev.\ Lett.\ {\bf#2}, #3 (#4)}
     \def\nuclph#1#2#3#4{#1, Nucl.\ Phys.\ B {\bf#2}, #3 (#4)}
     \def\nim#1#2#3#4{#1, Nucl.\ Instrum.\ Methods\ {\bf#2}, #3 (#4)}
     \def\zph#1#2#3#4{#1, Z. Phys.\ {\bf#2}, #3 (#4)}
     \def\zpc#1#2#3#4{#1, Z. Phys.\ C {\bf#2}, #3 (#4)}
     \def\etal{{\em et al.}}


\begin{thebibliography}{99}
\bibitem{footnote}We discuss $D^0$ decays explicitly in the text,
its charge conjugate decays are also implied
throughout the text unless otherwise stated.
\bibitem{early}
In the early days when people searched for
$D^0\bar{D}^0$ mixing using the hadronic method,
the lack of knowledge on DCSD decays and
the lack of statistics and decay time information have forced one to
assume that DCSD is negligible
in order to set upper limit for
mixing~\cite{SPEAR1,SPEAR2,E87,ACCMOR,DELCO,HRS,ARGUS,CLEO15}.
\end{thebibliography}
\end{document}